\documentclass[12pt,preprint]{aastex}

\newcommand{\dg}{$^{\circ}$}
\newcommand{\myemail}{conor.a.nixon@nasa.gov}
\newcommand{\water}{H$_2$O}
\newcommand{\methane}{CH$_4$}
\newcommand{\dmethane}{CH$_3$D}

\newcommand{\ethylene}{C$_2$H$_4$}
\newcommand{\ethane}{C$_2$H$_6$}
\newcommand{\propane}{C$_3$H$_8$}
\newcommand{\acet}{C$_2$H$_2$}
\newcommand{\diacet}{C$_4$H$_2$}
\newcommand{\methacet}{C$_3$H$_4$}
\newcommand{\cyanoacet}{HC$_3$N}
\newcommand{\coo}{CO$_2$}
\newcommand{\benzene}{C$_6$H$_6$}
\newcommand{\cm}{cm$^{-1}$}

\slugcomment{Submitted to Pub. Astron. Soc. Pacific}

\shorttitle{Titan Science with the JWST}
\shortauthors{Nixon et al.}

\begin{document}


\title{Titan Science with the James Webb Space Telescope (JWST)}


\author{Conor A. Nixon\altaffilmark{1}, Richard K. Achterberg\altaffilmark{2,1}, M{\'a}t{\'e} {\'A}d{\'a}mkovics\altaffilmark{3}, Bruno B{\'e}zard\altaffilmark{4}, Gordon L. Bjoraker\altaffilmark{1}, Thomas Cornet\altaffilmark{5}, Alexander G. Hayes\altaffilmark{6}, Emmanuel Lellouch\altaffilmark{4}, Mark T. Lemmon\altaffilmark{7}, Manuel L{\'o}pez-Puertas\altaffilmark{8}, S{\'e}bastien Rodriguez\altaffilmark{9}, Christophe Sotin\altaffilmark{10}, Nicholas A. Teanby\altaffilmark{11}, Elizabeth P. Turtle\altaffilmark{12}, Robert A. West\altaffilmark{10}}


\altaffiltext{1}{Planetary Systems Laboratory, NASA Goddard Space Flight Center, Greenbelt, MD 20771, USA. \myemail}
\altaffiltext{2}{Department of Astronomy, University of Maryland, College Park, MD 20742, USA.}
\altaffiltext{3}{Astronomy Department, University of California Berkeley, CA 94720, USA.}
\altaffiltext{4}{LESIA, Observatoire de Paris, CNRS, 92195 Meudon, France.}
\altaffiltext{5}{ESA ESAC, P.O. Box, 78 E-28691 Villanueva de la Ca{\~n}ada, Madrid, Spain.}
\altaffiltext{6}{Department of Astronomy, Cornell University, Space Science Building, Ithaca, NY 14853, USA.}
\altaffiltext{7}{Department of Atmospheric Sciences, Texas A\&M University, College Station, TX 77843, USA.}
\altaffiltext{8}{Instituto de Astrof{\'i}sica de Andaluc{\'i}a - CSIC, Glorieta de la Astronom{\'i}a, s/n. E-18008, Granada, Spain.}
\altaffiltext{9}{Laboratoire Astrophysique, Instrumentation et Mod{\'e}lisation (AIM), CNRS-UMR 7158, Universit{\'e} Paris-Diderot, CEA-Saclay, 91191 Gif sur Yvette, France.}
\altaffiltext{10}{Jet Propulsion Laboratory, California Institute of Technology, Pasadena, CA 91109, USA.}
\altaffiltext{11}{School of Earth Sciences, University of Bristol, Wills Memorial Building, QueenÕs Road, Bristol BS8 1RJ, UK.}
\altaffiltext{12}{Johns Hopkins University Applied Physics Laboratory, Laurel, MD 20723, USA.}


\begin{abstract}
The James Webb Space Telescope (JWST), scheduled for launch in 2018, is the successor to the Hubble Space Telescope (HST) but with a significantly larger aperture (6.5 m) and advanced instrumentation focusing on infrared science (0.6--28.0 \micron ). In this paper we examine the potential for scientific investigation of Titan using JWST, primarily with three of the four instruments: NIRSpec, NIRCam and MIRI, noting that science with NIRISS will be complementary. Five core scientific themes are identified: (i) surface (ii) tropospheric clouds (iii) tropospheric gases (iv) stratospheric composition and (v) stratospheric hazes. We discuss each theme in depth, including the scientific purpose, capabilities and limitations of the instrument suite, and suggested observing schemes. We pay particular attention to saturation, which is a problem for all three instruments, but may be alleviated for NIRCam through use of selecting small sub-arrays of the detectors - sufficient to encompass Titan, but with significantly faster read-out times. We find that JWST has very significant potential for advancing Titan science, with a spectral resolution exceeding the Cassini instrument suite at near-infrared wavelengths, and a spatial resolution exceeding HST at the same wavelengths. In particular, JWST will be valuable for time-domain monitoring of Titan, given a five to ten year expected lifetime for the observatory, for example monitoring the seasonal appearance of clouds. JWST observations in the post-Cassini period will complement those of other large facilities such as HST, ALMA, SOFIA and next-generation ground-based telescopes (TMT, GMT, EELT).

\end{abstract}





\section{Introduction}

The James Webb Space Telescope (JWST), currently planned for launch in 2018, is the most ambitious large space observatory since the launch of the Hubble Space Telescope (HST) in 1990. JWST is the undertaking of an international partnership of NASA, ESA and the Canadian Space Agency. This paper is the result of an investigation commissioned by the JWST Solar System Working Group (SSWG) in early 2014 to gather input from the planetary science community regarding the detailed capabilities of the JWST for solar system science. This resulted in the formation of ten Science Focus Groups (SFGs) for specific discipline areas: Mars, comets, NEOs etc. The goals for each of the groups were to: 
(a) describe specific scientific questions that could be addressed using JWST data; 
(b) summarize observation scenarios and data products needed to address those questions; 
(c) examine JWST instrument and observatory performance in light of the above.

This paper reports the findings of the Titan Science Focus Group (SFG). The Titan SFG included a volunteer membership drawn from the international scientific community active in Titan research, from a variety of institutional types (government agencies, universities, other research organizations), nationalities, and career backgrounds. The Titan SFG identified five key areas of potential high impact for the JWST in Titan science: (i) surface features (ii) lower atmosphere clouds (iii) gaseous composition of the troposphere (iv) gaseous composition of the stratosphere (v) stratospheric hazes. 

In the following section of the paper we give a brief summary of the JWST observatory. We describe the instrumentation, and compare to other large facilities that have observed Titan. We next discuss the general capabilities and restrictions of the observatory for observing Titan. In Section \ref{sect:examples} are the detailed findings for each of the five scientific subtopics, and in Section \ref{sect:conc} we summarize our conclusions and recommendations to the SSWG. 


\section{The JWST Observatory}
\label{sect:observatory}

\subsection{Observing Titan with JWST}
\label{sect:observing}

\subsubsection{Observability of Titan}

\begin{figure}
\includegraphics[scale=0.3]{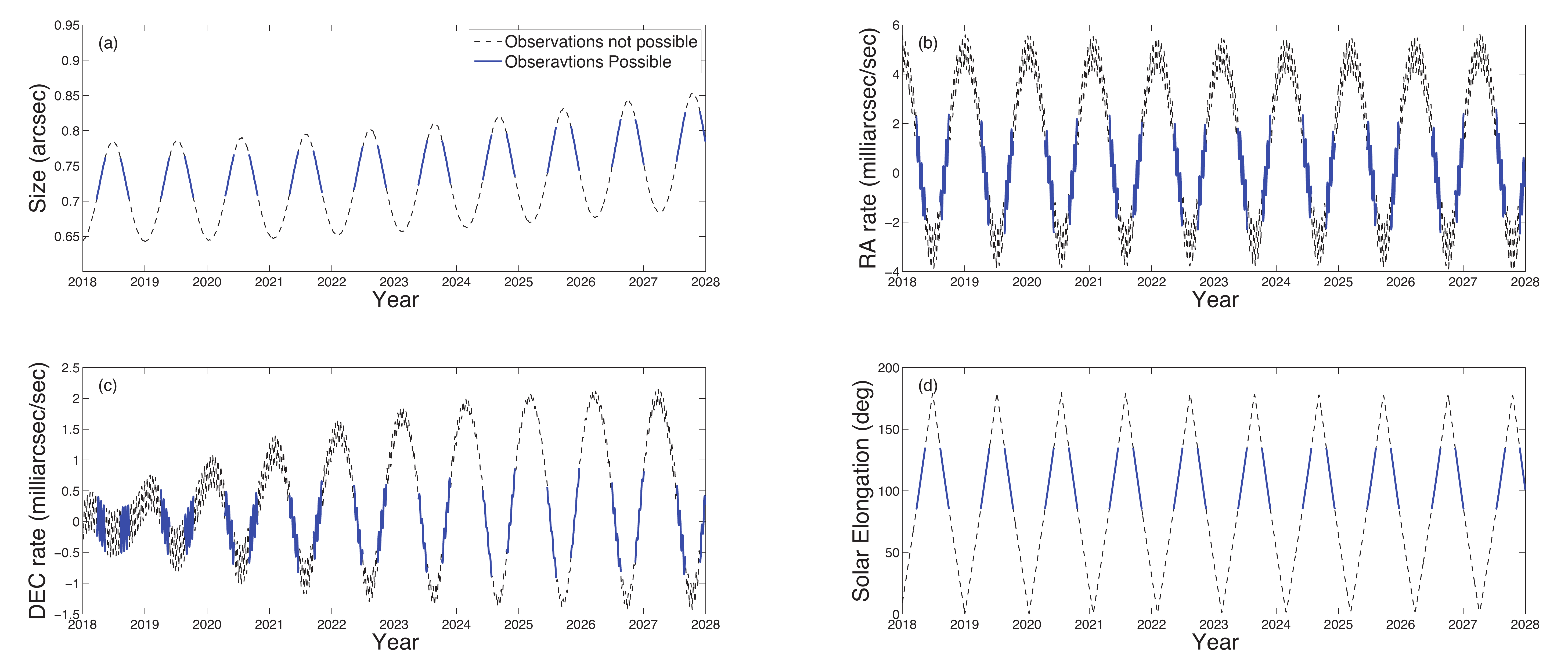}
\caption{Observability of Titan based on Solar elongation criteria for the JWST heat shield. There are gaps of up to half a terrestrial year where Titan is not observable (black parts of curves). This must be considered when planning seasonal time series observations. Ephemeris data is from JPL Horizons. JWST has a lifetime requirement of 5 years, but will carry fuel for $\sim$10 year mission.}
\label{fig:observability}
\end{figure}

The JWST was conceived as an infrared (0.6--28~\micron ) successor to HST, which operated primarily in the visible spectrum (0.1--2.5~\micron ), and therefore is intended to stay cold to enable sensitive measurements at long wavelengths. JWST will reside near the Earth-Sun Lagrange-2 (L2) point, where it will deploy a large sunshade to allow the telescope and instruments to reach an operating temperature of 50~K. Due to the shading requirements JWST is limited to observe at solar elongation angles between 85\dg\ and 135\dg , where the heat shield is effective, thereby limiting the cadence of repeat observations that are permitted (see Fig.~\ref{fig:observability}). Due to Titan's long seasons ($\sim$7.5 Earth years) there is no prospect of missing an entire season. Nevertheless, abrupt changes in global circulation are possible at much shorter timescales \citep[six Earth months, or 0.015 Titan years,][]{teanby12}. For example, near the equinox the observability of Titan may limit the ability of JWST to temporally resolve short-term seasonal changes. 

{\bf Figure~\ref{fig:observability} here}

\subsubsection{Resolution, pointing and tracking}

\begin{figure}
\includegraphics[scale=0.70]{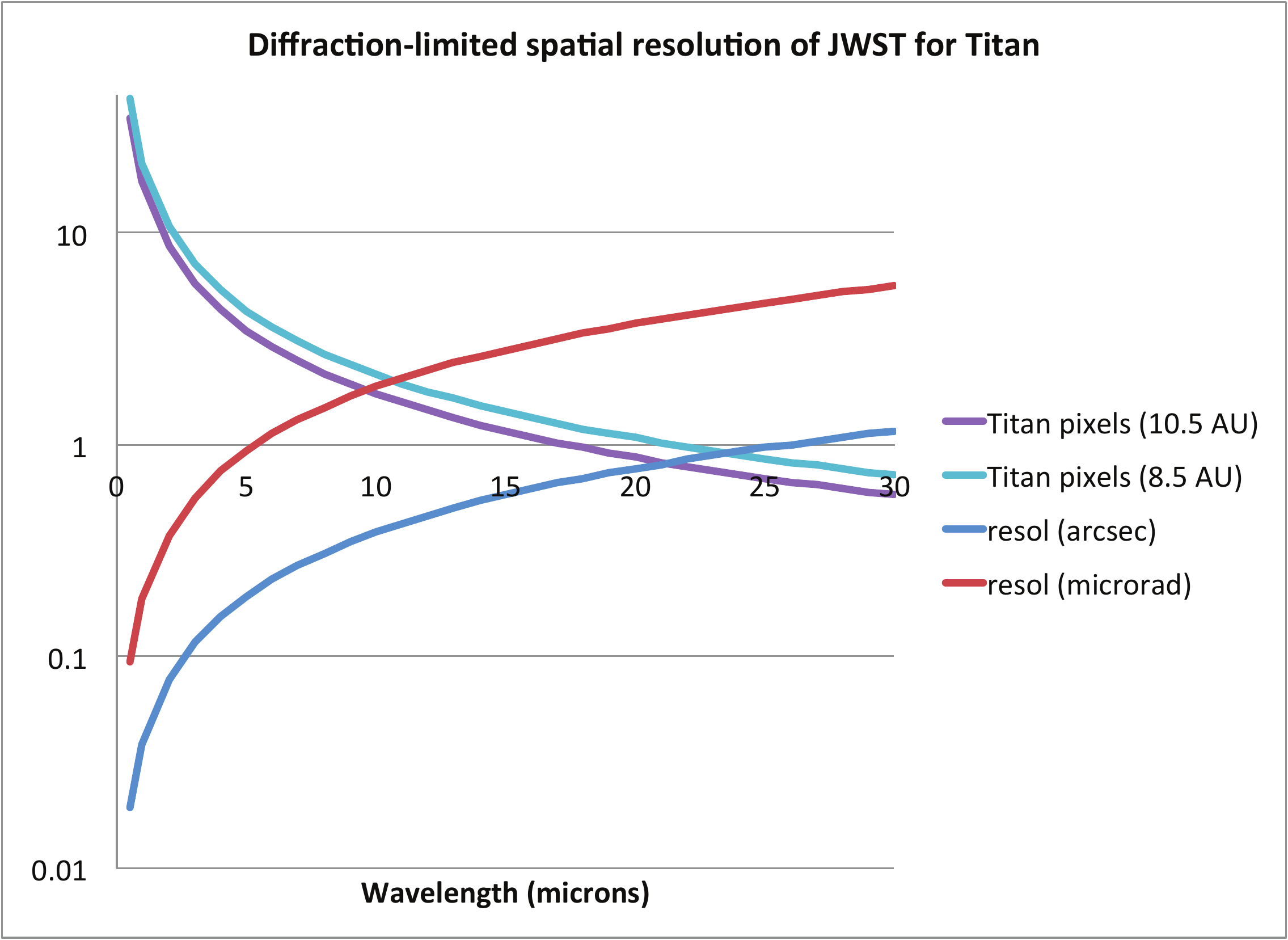}
\caption{Diffraction-limited spatial resolution of JWST (Airy disk radius) for Titan based on the primary 6.5~m aperture as a function of wavelength. The resolution is given both as the angular spot size (in microradians and arcsec) and the corresponding number of `pixels' (PSF resolution elements) across Titan's disk.}
\label{fig:resolution}
\end{figure}

\begin{figure}
\rotatebox{-270}{\includegraphics[scale=0.8]{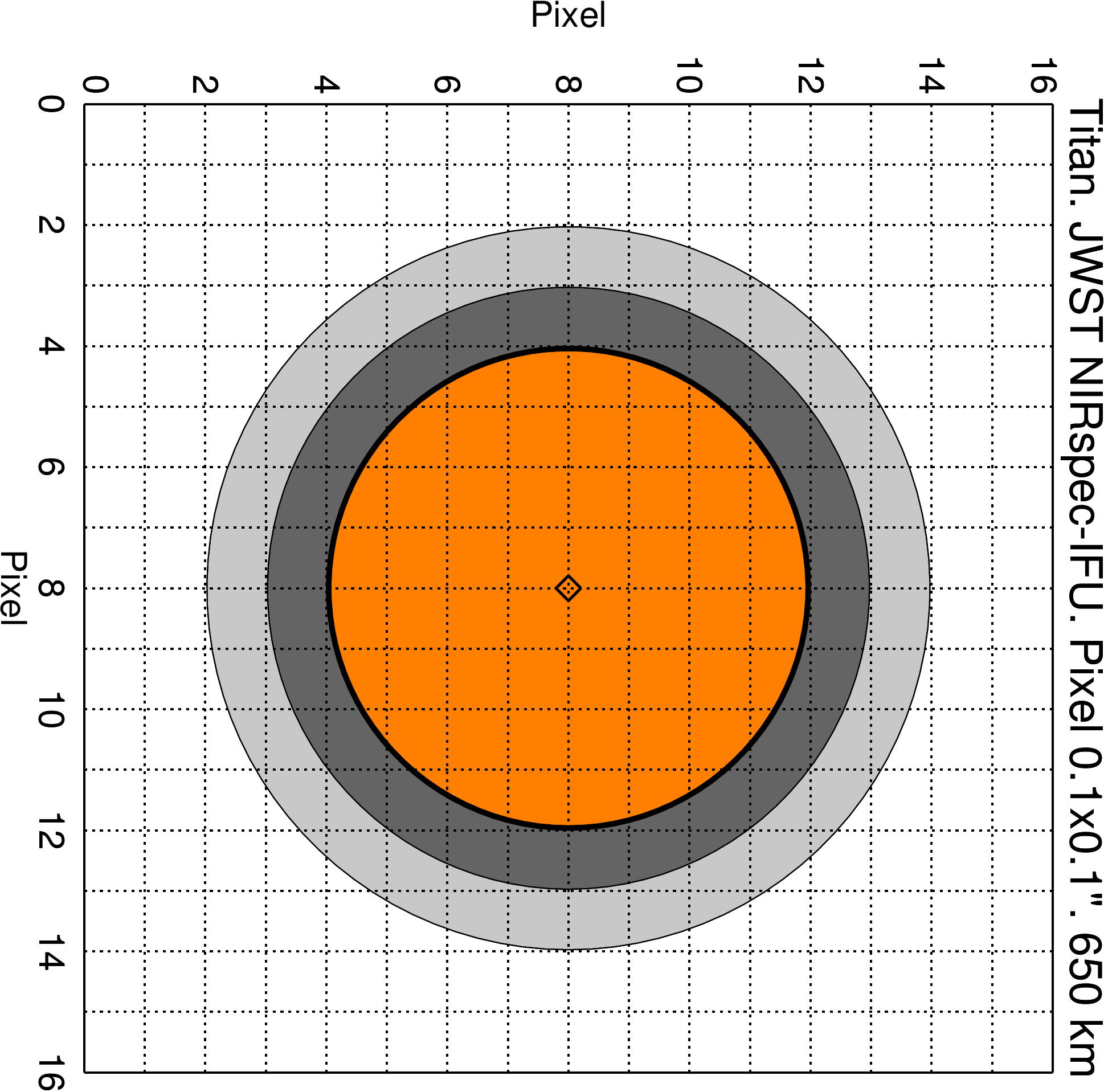} }
\caption{Expected spatial resolution on Titan (orange color represents solid body diameter) using NIRSpec IFU mode (0.1~{\arcsec}-square pixels, $\sim$650~km), showing that separation of the lower (dark grey) and upper (light grey) limb may be attained for bright fluorescent emissions.}
\label{fig:nirspecIFUresol}
\end{figure}

JWST has a primary aperture (6.5~m) that is more than twice the diameter of HST (2.4~m), but due to the longer wavelengths observed its resolution in the near-infrared is similar to Hubble's resolution in the visible. Figure~\ref{fig:resolution} shows the predicted spatial resolution for Titan, in terms of the telescope point spread function (PSF) radius (Airy radius, 1.22$\lambda/D$, where $\lambda$ is wavelength and $D=6.5$~m). Pixel angular sizes for the various instruments may restrict spatial resolution further (see Section~\ref{sect:instruments}). Note that Titan becomes spatially unresolved for wavelengths above $\sim$20 \micron . Figure~\ref{fig:nirspecIFUresol} shows a schematic of the achievable resolution in NIRSpec IFU mode: 0.1 {\arcsec}/pixel, corresponding to about 650~km.

{\bf Figure~\ref{fig:resolution} here}

{\bf Figure~\ref{fig:nirspecIFUresol} here} 

Note that JWST's maximum movement rate of 30 milliarcseconds/second (mas/s) is not limiting since Titan moves at most 6 mas/s on the sky. The pointing stability of 0.05{\arcsec}  is small with respect to Titan's diameter ($\sim$0.75{\arcsec}), although may have an impact on imaging at short wavelengths, where spatial resolution can reach 0.08{\arcsec} for NIRCam.

\subsection{JWST Instrumentation}
\label{sect:instruments}

\subsubsection{Overview of instrument suite}

The observatory is equipped with four instrument suites, each of which has multiple modes of operation: (i) NIRCam, a 0.6--5.0~\micron\ imaging camera; (ii) NIRSpec, a near-infrared spectrograph operating between 0.6 and 5.0~\micron ; (iii) MIRI, a 5.0--28.0~\micron\ mid-infrared spectrometer; (iv) NIRISS, primarily a guidance camera for JWST, but also capable of spectrometry, and operating at 0.8--5.0~\micron . Individually and in co-operation, these instruments provide a powerful and versatile means of monitoring the atmosphere and surface of of Titan. The near-infrared instruments (NIRCam, NIRISS, NIRSpec) view Titan primarily in reflected sunlight, including both opaque, atmospheric methane bands and semi-transparent spectral `windows' where the surface may imaged. MIRI on the other hand is uniquely capable on JWST at longer wavelengths, where thermal emission from Titan's cold ($\sim$75-150~K) atmosphere is measured. The relevant technical specifications of each instrument are now considered in more detail (see also Table~\ref{tab:specs}).

{\bf Table 1 here}

NIRCam is primarily an imaging camera, although with additional capabilities for spectroscopy and coronography that are not described here. The imaging system is composed of two independent systems: the short wavelength (SW, 0.6--2.3~\micron ) and long wavelength (LW, 2.4--5.0~\micron ) arrays. The SW system has four $2048\times 2048$ pixel detector arrays, for a total of 4096$\times$4096. The pixel pitch is 0.0317\arcsec , allowing for Nyquist sampling of telescope Point Spread Function (PSF, i.e. the Airy diffraction disk) at 2~\micron\ (see Fig.~\ref{fig:resolution}). For a Nyquist PSF size of 0.079{\arcsec} (2.5 pixels), there are 9.4 resolution elements across Titan's 5150~km solid-body diameter, each resolving 550~km. Including a 1000~km deep atmospheric limb takes the total diameter to 13 PSFs (32 pixels). The LW channel has just a single 2048$\times$2048 array, pixel pitch of 0.0648\arcsec , Nyquist-sampled PSF of 0.162{\arcsec} at 4 \micron , and 4.6 PSFs (15 pixels) across the disk resolving 1150~km each. NIRCam has numerous narrow, medium and wide band filters, described in more detail in Section~\ref{sect:clouds}. Sub-arraying (with consequent shorter read-out times) is required to avoid saturation in some filters (see Figs.~\ref{fig:nircamsat} and \ref{fig:nircamimag}). A subarray size of 160$^2$ pixels (both SW and LW) appears more than adequate to encompass Titan's disk plus atmosphere, while  reducing the read-out time by a factor of 38.8.

NIRSpec is a highly capable spectrometer and hyper-spectral imager, operating at near-infrared wavelengths, with three primary modes: (i) Micro-Shutter Array (MSA) mode, where a spectrum is obtained of a hundred or more point sources in the large field of view (3.4\arcmin $\times$3.6\arcmin , selected by opening and closing micro-shutter array doors); (ii) slit spectroscopy, used primarily for point sources, and (iii) Integral Field Unit (IFU) spectro-imaging, ideal for extended sources such as Titan. The IFU mode has a 3\arcsec $\times$3{\arcsec} FOV, sliced into 30 vertical slices of width 0.1\arcsec , and dispersed at a resolving power of 100, 1000 or 2700. The detector arrays are two 2048$\times$2048 pixel arrays. The IFU mode pixel sizes are 0.1\arcsec $\times$0.1{\arcsec}. Sub-arraying is possible, but probably not for the IFU mode where the full detector arrays are need for the hyper-spectral image, and so saturation is a problem for a relatively bright source like Titan (see Section \ref{sect:surface}).

MIRI is a spectro-imaging instrument operating in the mid-infrared. MIRI is capable of imaging, low-resolution slit spectroscopy, coronography, and - most important for extended sources such as Titan - an IFU mode (image slicer) known as the Medium Resolution Spectrograph (MRS). The MRS is divided into four wavelength ranges, each having a unique FOV size and spatial resolution: (1A) 4.96-7.71 \micron , (1B) 7.71-11.90 \micron , (2A) 11.90-18.35 \micron , and (2B) 18.35-28.30 \micron . The fields of view increase from 3.6{\arcsec}$\times$3.6{\arcsec} (1A) to 7.6{\arcsec}$\times$7.6{\arcsec} (2B). The rectangular pixel sizes for each range are: 0.18{\arcsec}$\times$0.19{\arcsec} (1A); 0.28{\arcsec}$\times$0.19{\arcsec}(1B); 0.39{\arcsec}$\times$0.24{\arcsec} (2A); 0.64{\arcsec}$\times$0.27{\arcsec} (2B). The spectral resolving powers are: 3250, 2650, 2000, 1550 respectively. 

The NIRISS instrument is also capable of spectroscopy of extended ($R\sim$150, 1.0--2.5~\micron ) and single sources ($R\sim$700, 0.6--3.0~\micron ), and imaging (1.0--5.0~\micron, 2.2{\arcmin}$\times$2.2{\arcmin} field). NIRISS images onto a 2048$\times$2048 pixel array with pitch 0.0654 {\arcsec}/pixel and can obtain `blue' and `red' images simultaneously, an advantage over NIRCam.

\subsubsection{JWST instruments compared to other instruments}

In this paper we will compare JWST Titan science capability to the science already achieved by other facilities and instruments, in particular: NIRSpec to the Cassini Visual and Infrared Mapping Spectrometer \citep[VIMS,][]{brown04}; NIRCam to the Cassini Imaging Science Subsystem \citep[ISS,][]{porco04} and HST; MIRI to the Cassini Composite Infrared Spectrometer \citep[CIRS,][]{flasar04b}. The spatial resolution of HST and JWST are similar, however the Cassini instruments are in general far superior to JWST in terms of spatial resolution, by virtue of the orbiter's up-close encounters with Titan. ISS has imaged $\sim 90$\% of Titan's surface at resolution better than 5~km, and $\sim$50\% at a few km resolution, down to the limit of $\sim$1~km imposed by atmospheric scattering \citep{porco04, stephan10b}. VIMS maps of Titan's surface have been constructed using a requirement of $<$100~km/pixel, but at lower latitudes resolutions of 0.5--2.0~km have been achieved under optimal conditions \citep{stephan10b, lemouelic12a}. CIRS maps of Titan's surface have a much lower resolution, at best $\sim$300~km, due to the large detector footprint at 19~\micron\ where the surface is sensed \citep{cottini12a}. However at this wavelength JWST/MIRI will not resolve Titan. CIRS also produces detailed vertical profiles of gases by limb mapping at resolutions of one scale height, 50~km during close flybys. Cassini has one further advantage over JWST, which is the ability to view Titan at all (including high) phase angles, enabling additional information gathering.

\begin{figure}
\epsscale{0.70}
\plotone{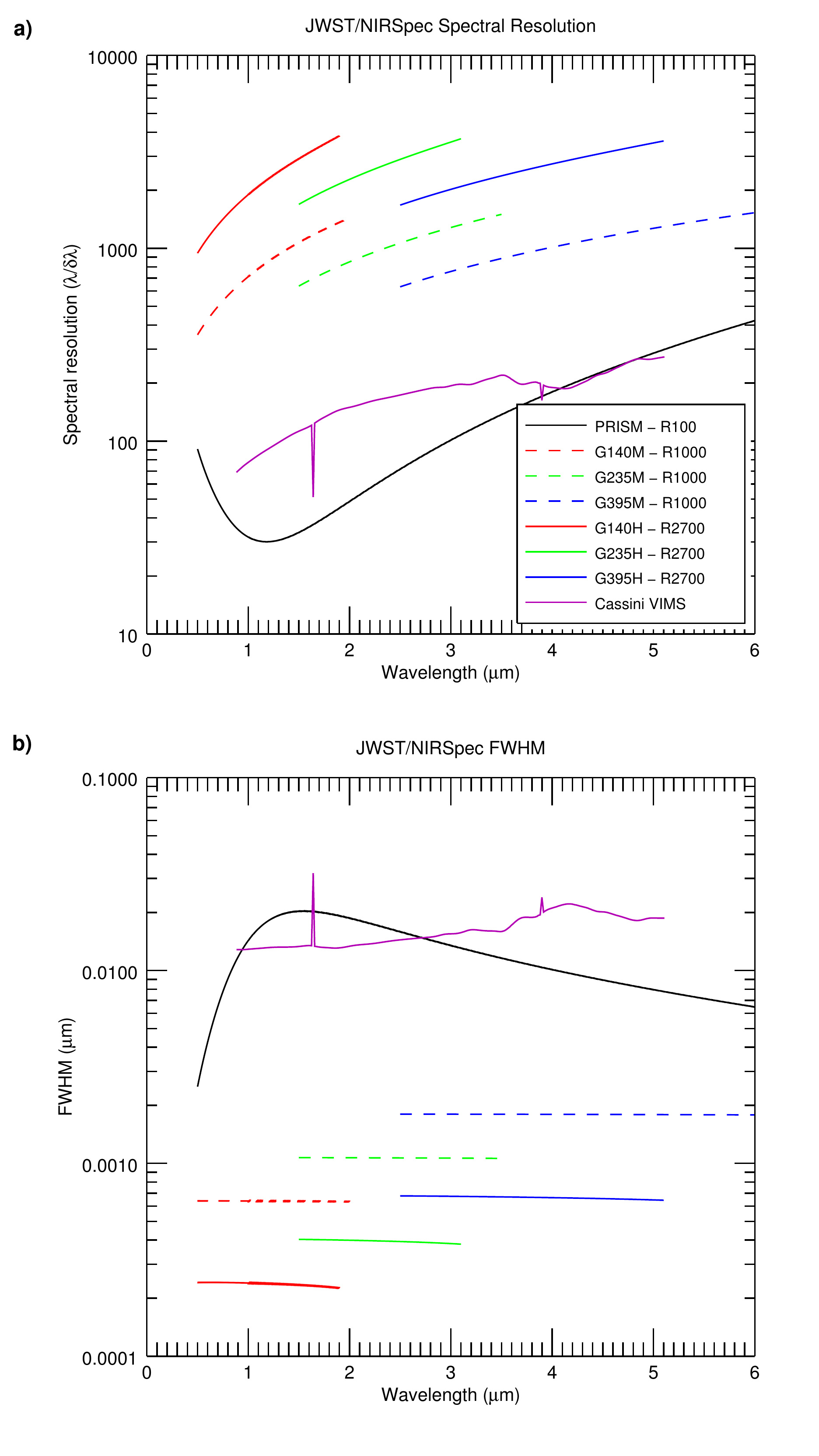}
\caption{NIRSpec spectral resolution (top) and FWHM (bottom) compared to Cassini VIMS. In the legend, PRISM indicates the use of a single prism to collect the spectra, while G\#\#\#M/H indicate the use of gratings with a peak efficiency approximately centered at \#.\#\# microns at high (R=2700) or medium (R=1000) resolution.}
\label{fig:vimsnirspec}
\end{figure}

Spectrally, the comparison between JWST and Cassini is more favorable: the highest spectral resolution of MIRI ($R\sim 2800$) is similar to that of CIRS at 7~\micron\ (shortest wavelength of CIRS), and some $2\times$ higher at 28~\micron\ (longest wavelength of MIRI). NIRSpec can achieve a much higher spectral resolution than VIMS (see Fig.~\ref{fig:vimsnirspec}). VIMS has a resolving power $R$ varying from 70 to 270 across the spectral range. NIRSpec has three resolutions:  R100 using a single prism, yielding $R$ = 30--300, similar to VIMS; and R1000 and R2700 each using three gratings, yielding $R$ = 300--1200 and $R$ = 1000--2700 respectively. Note that NIRSpec's Full Width at Half Maximum (FWHM) is nearly constant for R1000 and R2700 modes, but variable for R100 (R=30--300). It is beyond the scope of this paper to describe the capabilities of the various HST instruments that have been used to observe Titan, however it is worth pointing out that the longest observable wavelength is 2.2~\micron\ (NICMOS), so the science permitted is quite different from JWST and Cassini.

{\bf Figure \ref{fig:vimsnirspec} here}


\section{Example Science Investigations}
\label{sect:examples}

In this section we describe example science investigations organized into five thematic areas, and focusing on the science achievable with three of the four JWST instruments: NIRSpec, NIRCam and MIRI. The capabilities of the fourth instrument, NIRISS, significantly overlap with the other instruments, and therefore the science investigations permitted are similar and not separately described. While we intend to have covered the most obvious and important topics, this list of investigations is surely incomplete, and during the lifetime of JWST we fully expect that this list will be substantially expanded.      

\subsection{Titan's surface features}
\label{sect:surface}

{\em Scope: near-infrared spectroscopy and imaging of Titan's surface to determine features, composition, geology, history. Instruments: NIRCam, NIRSpec, MIRI.}

The surface of Titan is a veneer of photochemical products derived from nitrogen and methane photolysis and whose exact nature is still unknown, deposited on a water ice substrate. Under present day surface conditions, methane, ethane, and propane are (meta)stable in the form of liquids. Fluvial \citep{burr13b}, aeolian \citep{lucas14a}, and lacustrine \citep{mackenzie14} processes are known to alter the landscape and there may be more dramatic alterations due to impacts \citep{neish12} and volcanism \citep{lopes13}. The presence of a rich mixture of organic material in contact with a large reservoir of water is among the motivations for further exploration. Beginning with HST WFPC-2 observations in the 1990s \citep{smith96}, the composition and physical state of the surface have been studied using near-infrared imaging and spectroscopy at wavelengths that can penetrate through the atmosphere (known as spectral `windows') between methane absorption bands.  The dominant science questions remain: What is the exact surface composition and texture? Is there any exposed water ice at the surface of Titan? What processes are changing the surface?
JWST can make critical contributions towards addressing these questions by observing the distribution of exposed ices (if any) on the surface, identifying and cataloging the hydrocarbons and nitrile materials, and monitoring the atmospheric, geological, and geophysical processes that alter the distribution of surface materials and structure of the surface itself. At 2--5 \micron, the spatial resolution of JWST is similar to that of HST imaging of Titan at $\sim1$ \micron, which was sufficient to resolve the largest spatial terrains \citep[e.g. the bright, equatorial Xanadu albedo feature,][]{smith96}.

\subsubsection{Variation in exposed water ice}
Ground-based observations of Titan's surface are challenging because of the limitations imposed by having to look through Earth's water and oxygen-rich atmosphere and through Titan's methane-rich atmosphere. JWST will solve half this problem by virtue of its location at L2, alleviating the need to remove telluric contamination to water ice absorption spectra and the associated photometric uncertainties. The combination of low spatial resolution (e.g. 8 resolution elements for NIRSpec IFU mode, Fig.~\ref{fig:nirspecIFUresol}) and photometric accuracy will strengthen the identification of individual spectral units and facilitate monitoring of surface changes due to resurfacing processes. Due to the opacity of TitanÕs atmosphere in the infrared, water ice cannot be directly identified from spectral absorption features. Nevertheless, specific inter-window spectral slopes can be positively correlated with a water ice signature, or at a minimum the signature of a possible enrichment in water ice content. For instance, low values of the 1.59/1.08~{\micron}, 1.59/1.27~{\micron}, 2.01/1.08~{\micron} and 2.01/1.27~{\micron} ratios relative to the 1.27/1.08~{\micron} ratio, all corrected from atmospheric haze scattering, indicate the presence of a surface constituent that absorbs more at 1.59 and 2.01~{\micron} than at 1.08 and 1.27~{\micron}, which is fully compatible with simulated and laboratory spectra of fine-grained water ice \citep{rodriguez06, rodriguez14}.

\subsubsection{Composition and distribution of organic material} 

\begin{figure}
\epsscale{0.80}
\plotone{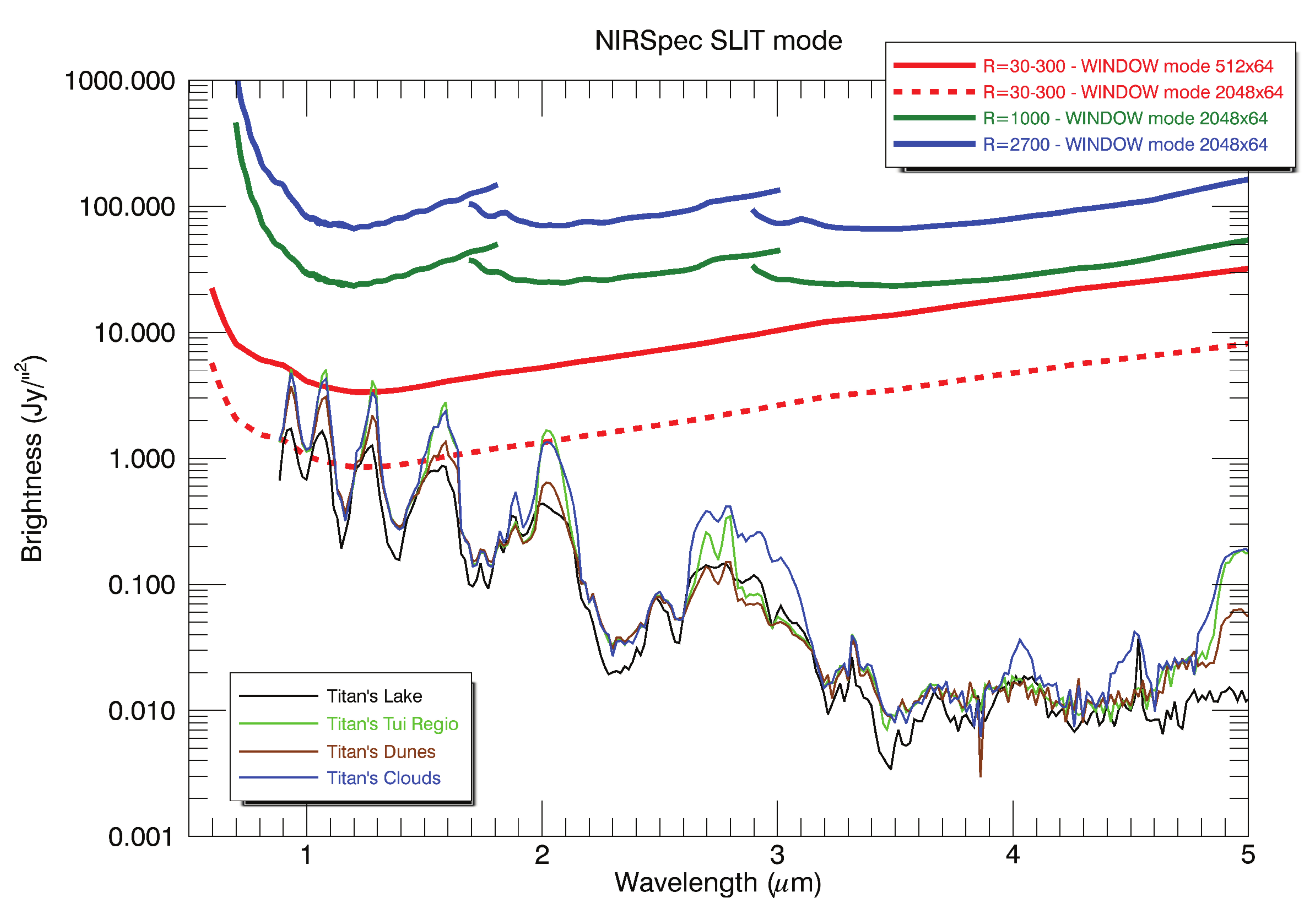}
\plotone{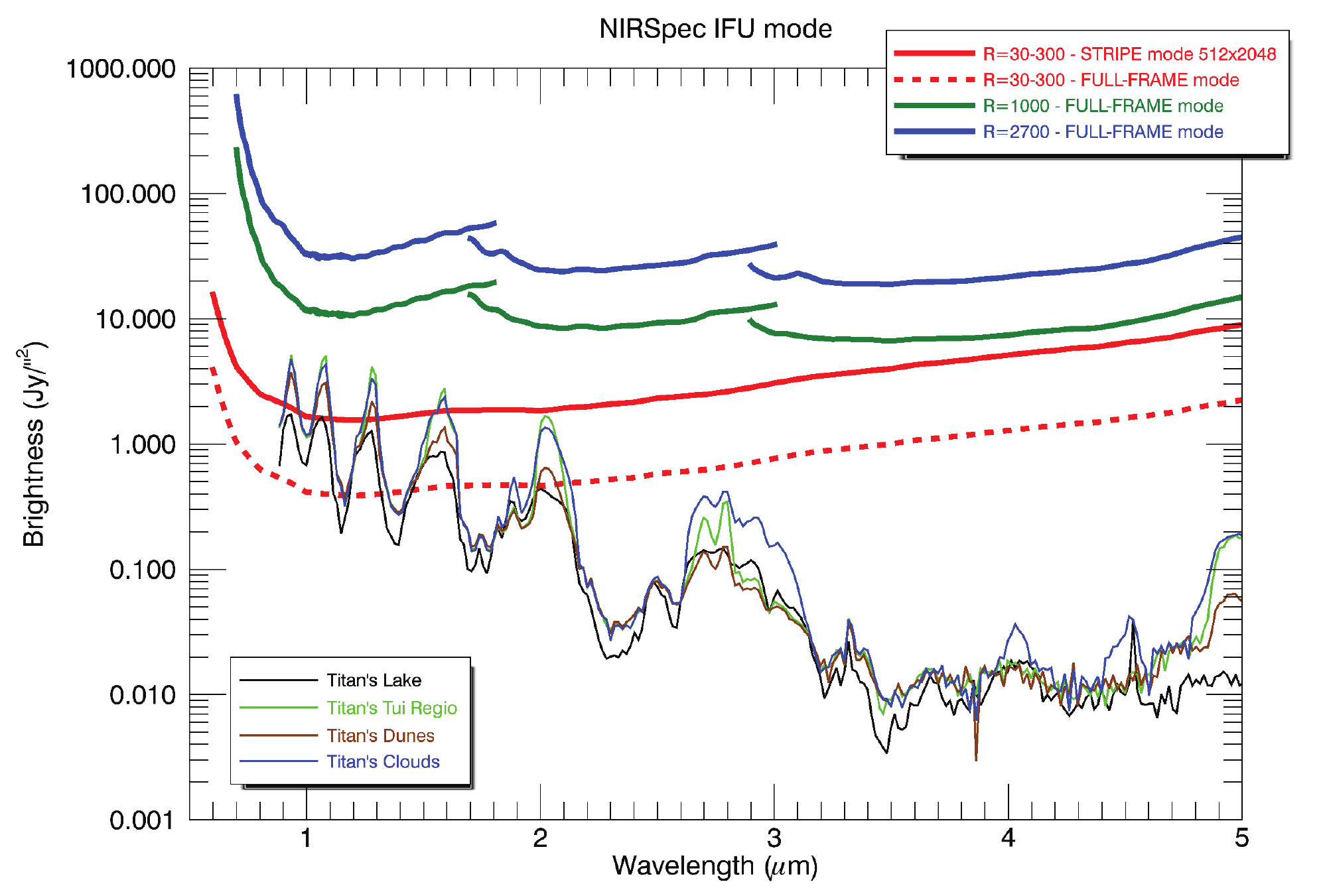}
\caption{Titan NIR spectrum for various terrain types based on Cassini VIMS data. Shown above are the saturation thresholds for various slit (top) and IFU (bottom) observing modes of NIRSpec, showing that low spectral resolution modes will saturate the detectors around 1~\micron , and up to 2~\micron\ for standard frame read-out times.}
\label{fig:nirspecsat}
\end{figure}

{\bf Figure \ref{fig:nirspecsat} here}

Liquid and solid hydrocarbons and nitriles are produced by Titan's complex atmospheric photochemistry, and range from relatively simple hydrocarbons (e.g. ethane) to highly complex C-H-N compounds \citep{atreya06}. These exhibit many absorption bands in the 1--5 \micron\ wavelength range \citep{clark10} accessible to NIRSpec that are blended at the low resolving power (R$\sim$150) of Cassini VIMS (Fig. \ref{fig:nirspecsat}), but will show up strikingly in NIRSpec starting at $R\sim 500$ and better S/N (Fig. \ref{fig:ethane}). Most of these bands have typical widths of 0.02--0.10 \micron\ in the 4.8--5.1 \micron\ range fully accessible to NIRSpec, allowing for possible new identifications, and mapping of large-scale terrain types with NIRCam of compounds at a spatial sampling close to that of HST. In the worst case (PRISM mode), the spectral resolution of NIRSpec will be similar to VIMS (see Fig.~\ref{fig:vimsnirspec}), and in the best case, its resolution will be similar to that of laboratory spectra. However, care will need to be taken to avoid saturation of NIRSpec at the shortest wavelengths. This applies equally to observations directed toward surface science or clouds (Fig.~\ref{fig:nirspecsat}). Typically, observations using the PRISM mode of NIRSpec must be avoided because of such saturation problems. However, observations using the IFU mode of NIRSpec will be possible with a resolving power of 1000 or more and a single exposure (about 12 seconds per frame, already much longer than is usually done with VIMS), for which sharp absorptions appear in the spectrum. For example at high spectral resolution the 0.7--5.2~\micron\ range can be covered in four exposures using specific filter/grating settings: F070LP/G140H, F100LP/G140H, F170LP/G235H, F290LP/G395H. Multiple exposures could be used to increase the S/N for weak features, but shorter exposures to avoid saturation may not be possible for IFU mode where the entire pixel array is needed.

\begin{figure}
\includegraphics[scale=0.4]{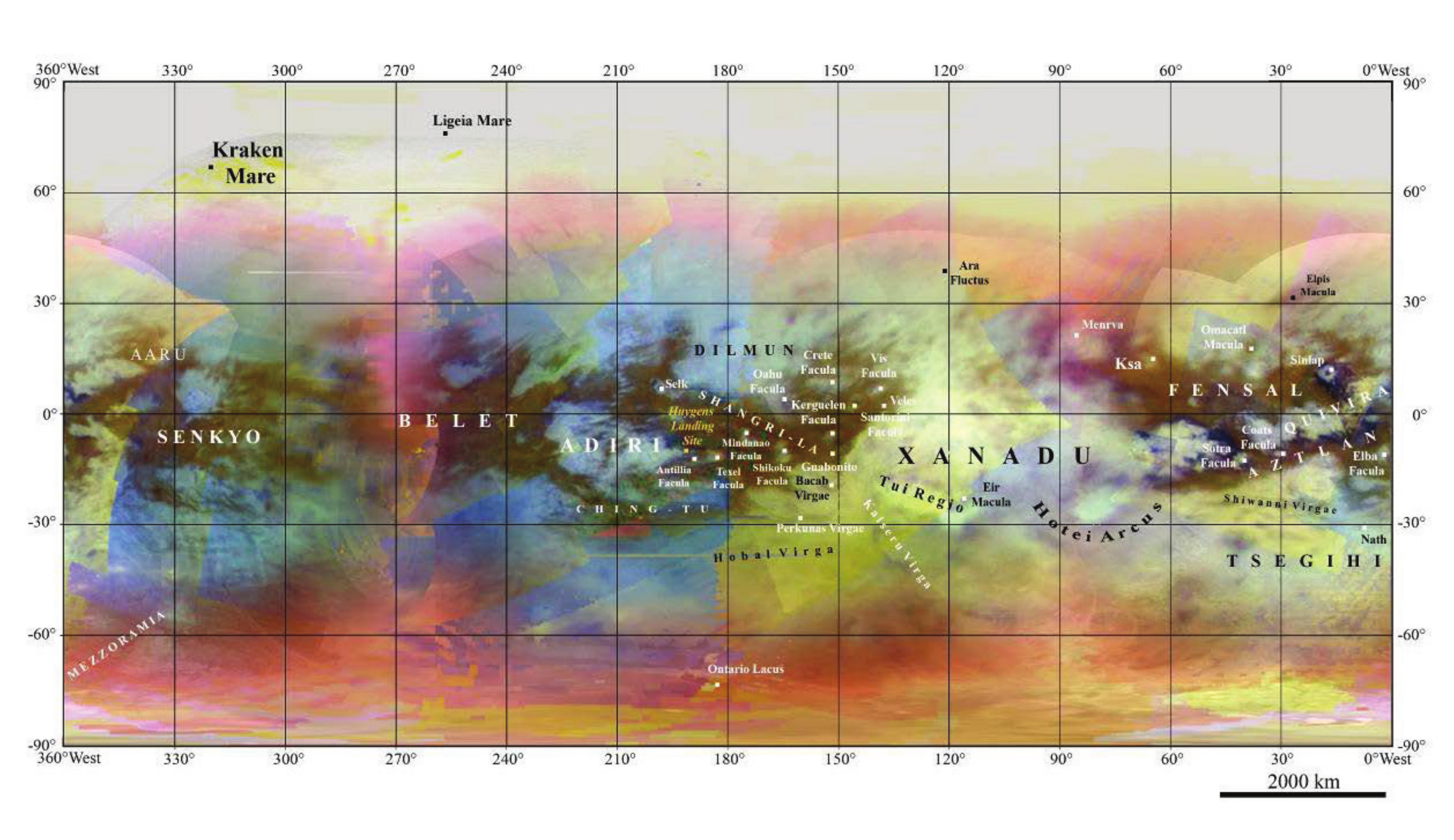}
\caption{Global VIMS cylindrical-projection map of Titan's surface. This false-color image uses: Red = 1.59 /1.27 \micron , Green= 2.03 / 1.27 \micron , Blue = 1.27 /1.08 \micron\ (Europlanet IDIS/K. Stephan).}
\label{fig:vimsmap}
\end{figure}

\begin{figure}
\includegraphics[scale=0.35]{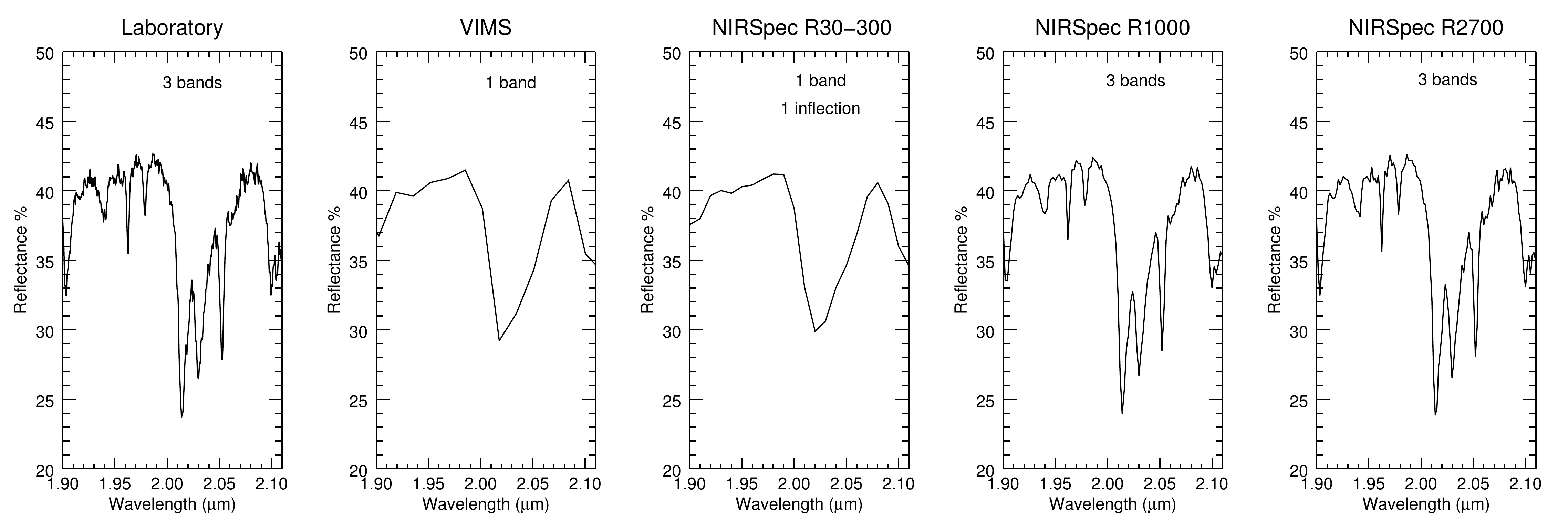}
\caption{Comparison of laboratory spectrum of liquid ethane at 2.0~\micron , with the same absorption viewed at VIMS resolution, and three progressively increasing resolutions available with the JWST NIRSpec instrument. The triple band at 2~\micron\ shows up as soon as the resolution approaches 1000. Laboratory spectra from the Arkansas Center for Space and Planetary Sciences.}
\label{fig:ethane}
\end{figure}

Note that the spatial resolution achievable with JWST at near-infrared wavelengths will not remotely rival that achieved by Cassini, with its vantage point in Saturn orbit. For  near-IR imaging Cassini has achieved 1~km (ISS), versus 200~km (JWST NIRCam); for near-IR spectroscopy the resolution is $\sim1$~km (Cassini VIMS) compared to 400~km (JWST NIRSpec), and for mid-IR spectroscopy 200~km (Cassini CIRS) versus 2000~km (JWST MIRI). However, these differences may be seen as complementary. Existing near-infrared Cassini maps can provide a higher resolution context for the JWST observations, and allow the targeted selection of large regions of particular interest still accessible to JWST spatial resolution: large polar seas (Kraken Mare, the largest sea, extends down to $\sim$55\dg N latitude); vast equatorial dune fields; equatorial bright regions, especially Xanadu, Tui Regio, and Hotei Regio; and mid-latitude `bland land' regions (Fig.~\ref{fig:vimsmap}).

{\bf Figure \ref{fig:vimsmap} here}

Hydrocarbons and nitriles also have absorption bands in the 2.75 and 2.00~\micron\ windows (i.e. solid and liquid ethane, benzene, cyanoacetylene), which can be used to validate some of the potential detections made at longer wavelengths. As an example, Fig.~\ref{fig:ethane} demonstrates how detection of liquid ethane, a likely ingredient of Titan's lakes/seas, in the 2.0~\micron\ window becomes progressively more certain as higher spectral resolving powers are used. More complex mixtures of hydrocarbons, nitriles, \water\ and tholin-like deposits can be also investigated using all the available NIRSpec/MIRI transparency windows, but will require the use of radiative transfer modeling to remove atmospheric contribution.

{\bf Figure \ref{fig:ethane} here}

\subsubsection {Active geology}
\begin{figure}
\plotone{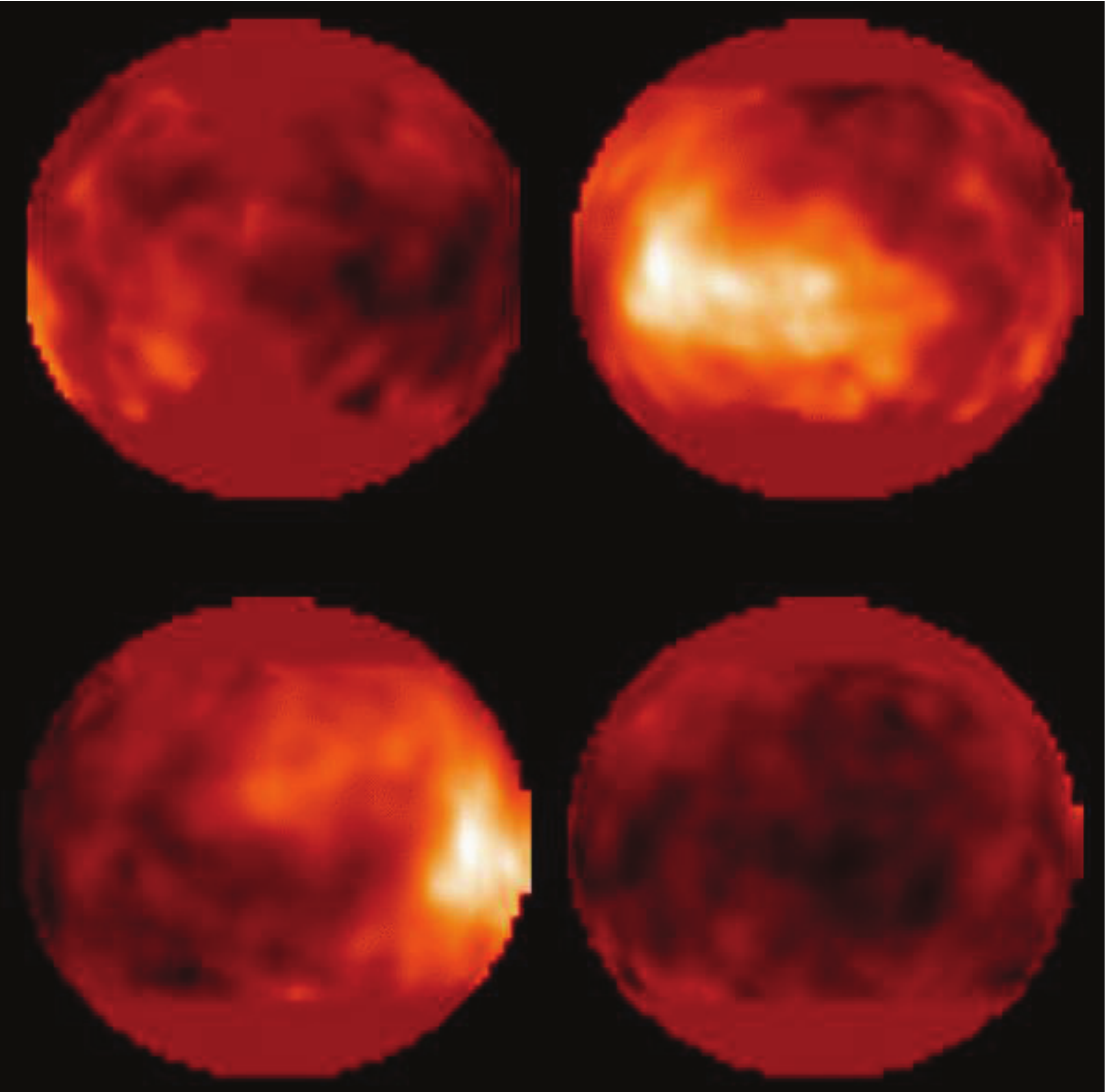}
\caption{First image showing Titan's surface, using the HST WFPC-2 F850LP filter to peer through the methane window at 940 nm \citep{smith96}. Note the bright Xanadu ÔcontinentÕ (on leading hemisphere, upper right), previously inferred from the rotational light curve, was imaged for the first time. Credit: NASA/JPL/STScI. Image PIA01465.}
\label{fig:titanhst}
\end{figure}

Advances in geomorphology are unlikely given the limited spatial resolution of JWST (see Fig.~\ref{fig:resolution}). However, at the largest scale, NIRCam - cross-compared with NIRSpec - could be used to measure the albedos of the bright/dark terrains, e.g. through rotational light curves \citep{buratti06}. JWST spatial resolution is comparable or better that than of the HST, which was able to image the bright Xanadu region on Titan's leading hemisphere (Fig.~\ref{fig:titanhst}). At the smallest scale, the same could be achieved for sufficiently large geomorphological units such as the large seas, dune fields, potentially cryovolcanic regions \citep[e.g.][]{lopes13}, poorly understood mid-latitudes (as identified with Cassini), as well as monitoring changes due to geologic or atmospheric activity (see Section~\ref{sect:clouds}). JWST resolution will not however be sufficient to monitor changes in sea shorelines, expected at scales of several km.

{\bf Figure \ref{fig:titanhst} here}

\subsubsection{Surface temperature}
	
By monitoring the spectral window at 19~\micron\ with MIRI \citep[as demonstrated by CIRS/Cassini measurements, e.g. ][]{jennings09a, jennings11, cottini12a}, it will be possible to measure Titan's disk-average (low-latitude) surface temperature and follow its evolution with seasons, giving a measure of thermal inertia. Also, although not observed to date, cryovolcanic temperature anomalies due to transient hotspots and/or extrusion of hot ice at 200--300~K would be detectable by MIRI Ð provided that the spatial extent is large enough.

\subsection{Clouds in Titan's lower atmosphere}
\label{sect:clouds}
{\em Scope: long-term monitoring campaign and/or target of opportunity (TOO) type quick-response observations of transient clouds and surface darkening. Instruments: NIRCam, NIRSpec.}

\subsubsection{Previous observations of Titan clouds:}
\begin{figure}
\plotone{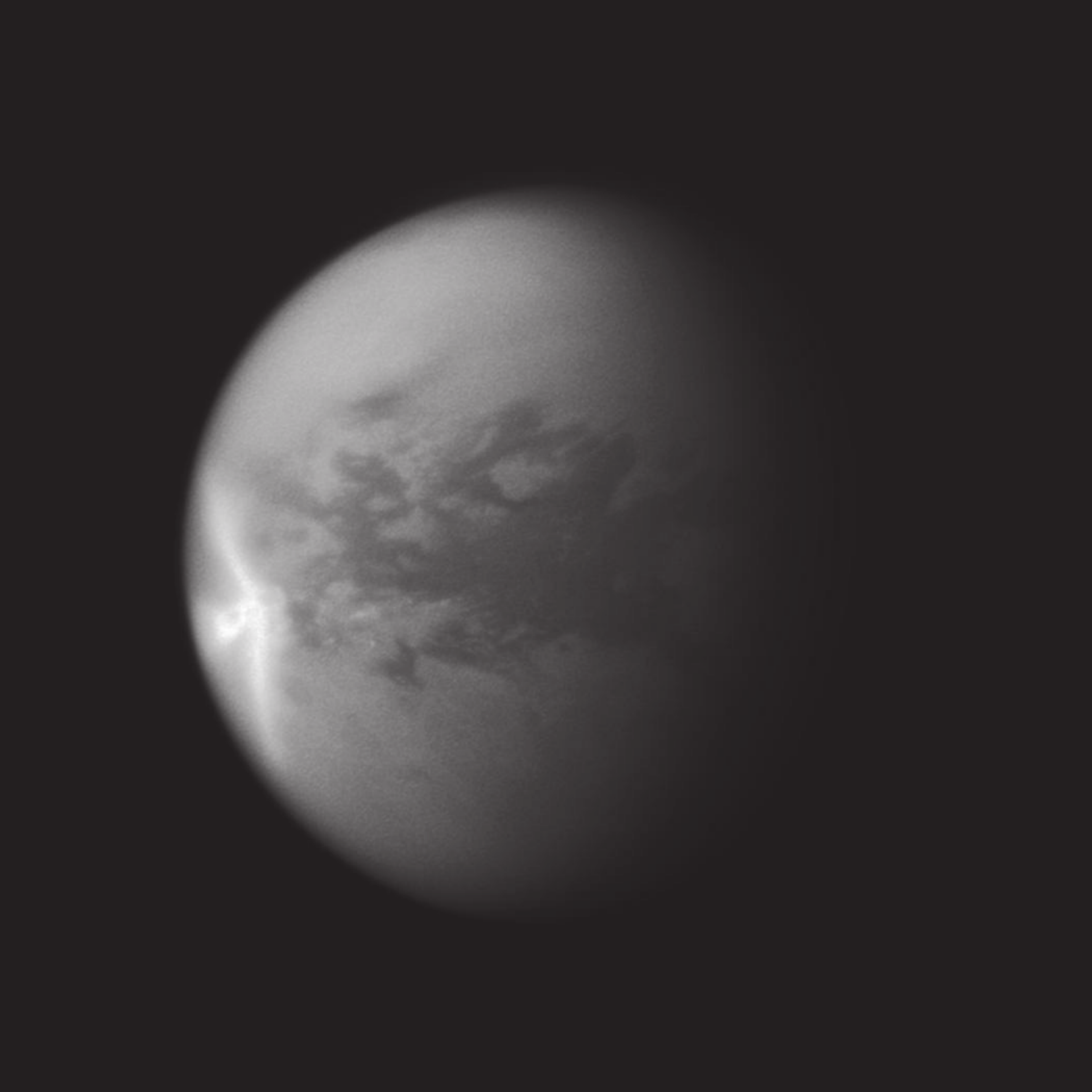}
\caption{A giant arrow-shaped cloud (bright feature on left side) was seen on Titan by the Cassini Imaging Science Subsystem (ISS) in March of 2011 \citep{turtle11b}. Credit: NASA/JPL/STScI. Image PIA12817.}
\label{fig:arrowcloud}
\end{figure}

Clouds were first detected spectroscopically on Titan in 1995 \citep{griffith98} and confirmed by direct imaging a few years later \citep{brown02}. Cassini has greatly expanded our understanding of cloud occurrence \citep{rodriguez09, rodriguez11, brown10, turtle11b}, revealing dramatic, unexpected morphologies such as the giant `arrow-shaped' cloud of 2011 (Fig.~\ref{fig:arrowcloud}). The storm - seen here stretching 1200~km E--W and 1500~km N--S - caused rainfall that temporarily darkened the surface coverage over 500,000~km$^2$, before the methane re-evaporated \citep{turtle11b}, followed by localized surface brightening \citep{barnes13b}. Such phenomena could be monitored by JWST, which can be targeted rapidly (48 hrs) using TOO (ÒTarget of OpportunityÓ) requests, and clearly has sufficient resolution at near-IR wavelengths with NIRCam and NIRSpec to resolve such large cloud features. Cloud structures may well occur down to the resolution limit of these instruments, and below, where JWST will not have the spatial resolution of Cassini to resolve fine detail.

{\bf Figure \ref{fig:arrowcloud} here}

\begin{figure}[ht]
\includegraphics[]{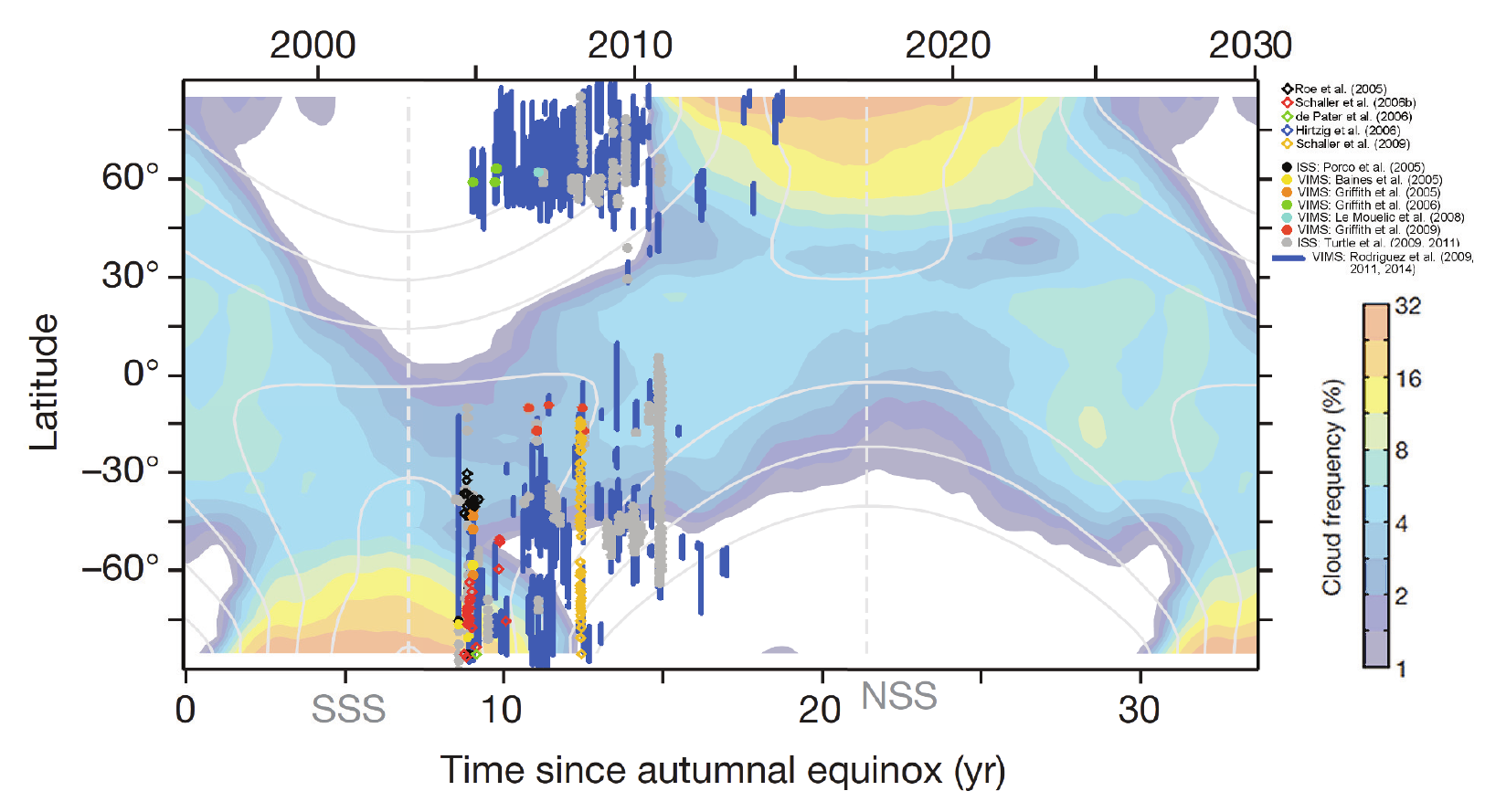}
\caption{Comparison of clouds observed with Cassini and ground-based telescopes to predictions of cloud frequency as a function of season \citep{schneider12}.}
\label{fig:cloudmodel}
\end{figure}

\begin{figure}[ht]
\plotone{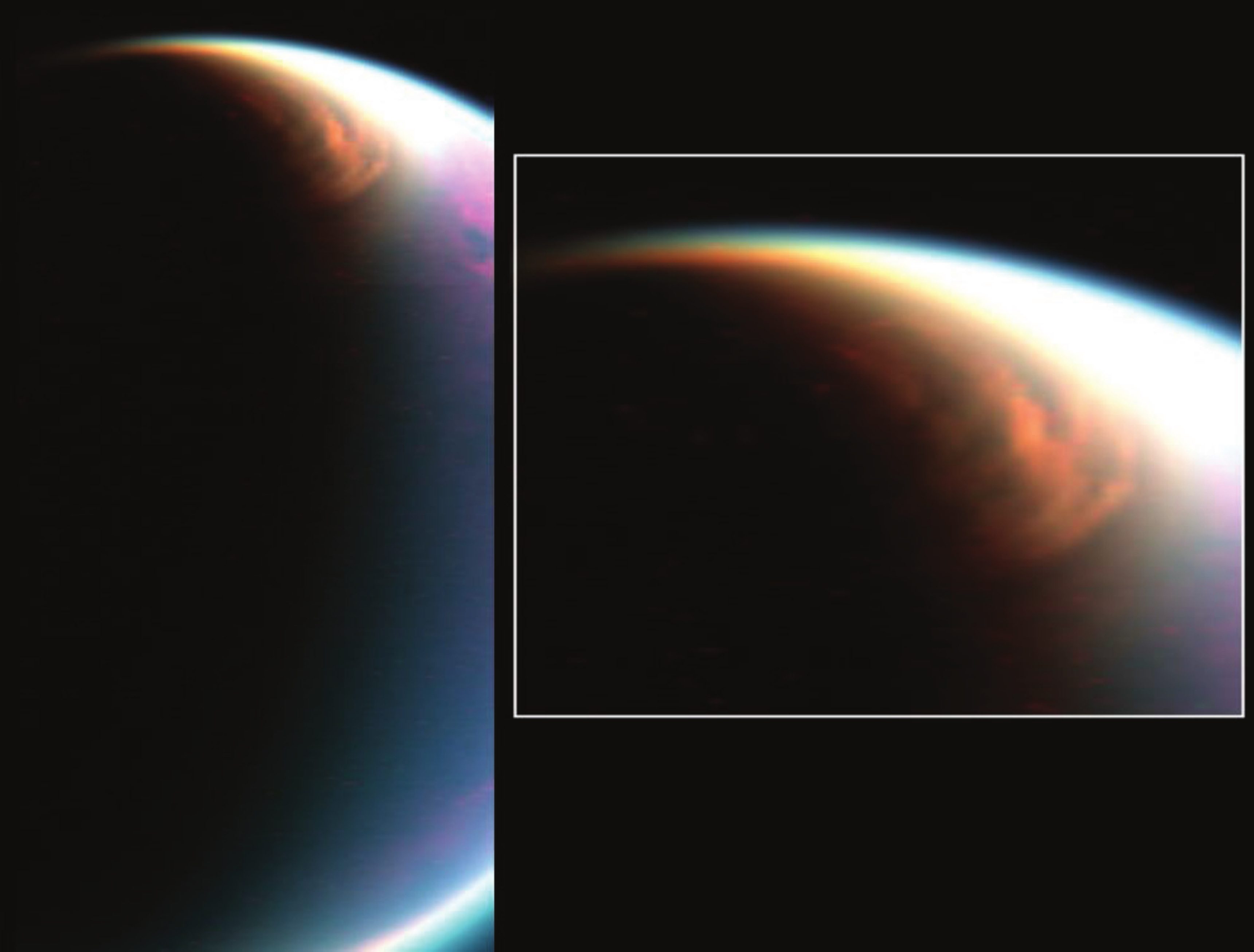} 
\caption{Giant high-altitude cloud covering Titan's north pole as seen by Cassini VIMS in 2007, and proposed to be composed of ethane. Color-coded: Red=2.0 \micron , Blue=2.7 \micron , Green=5.0 \micron . Image: PIA 09171. Credit: NASA/JPL/University of Arizona/LPGNantes.}
\label{fig:npolecloud}
\end{figure}

In addition, Cassini has proven a powerful tool for monitoring the latitudinal and temporal distributions of the clouds (Figs. \ref{fig:arrowcloud}--\ref{fig:npolecloud}) \citep{west15, rodriguez09, rodriguez11, brown10, turtle09, turtle11c, lemouelic12b}. The time-varying cloud coverage to date is compared to a seasonal prediction by \cite{schneider12} in Fig.~\ref{fig:cloudmodel}. 

{\bf Figure \ref{fig:cloudmodel} here}

{\bf Figure \ref{fig:npolecloud} here}

\subsubsection{Observing clouds with JWST}

Long-term monitoring is key to understanding Titan's weather patterns and atmospheric circulation. JWST NIRSpec will enable the detection of convective methane clouds and winter subsiding ethane clouds by simultaneously inspecting the shapes of the 2.75 and 5.00 \micron\ windows (Fig.~\ref{fig:nirspecsat}), with a higher sensitivity than Cassini. The additional absorption bands viewable due to lack of sky background will prove an asset to JWST observations compared to ground-based facilities. Mapping - in terms of percentage regional coverage - and documenting long-term evolution of cloud cover will also be feasible. Titan's clouds evolve on time scales of days \citep{turtle11b}, so short-term monitoring of evolution (shape, motion, opacity) over days to weeks will likely necessitate a request for a rapid response, event-driven (TOO) observation, with a 48 hour lag. The  improved spectral resolution of NIRSpec will permit better constraints on cloud physical properties (altitude, droplet size and composition) and possibly distinguish between methane and ethane clouds, which is challenging with Cassini.

\begin{figure}
\includegraphics[scale=1.4]{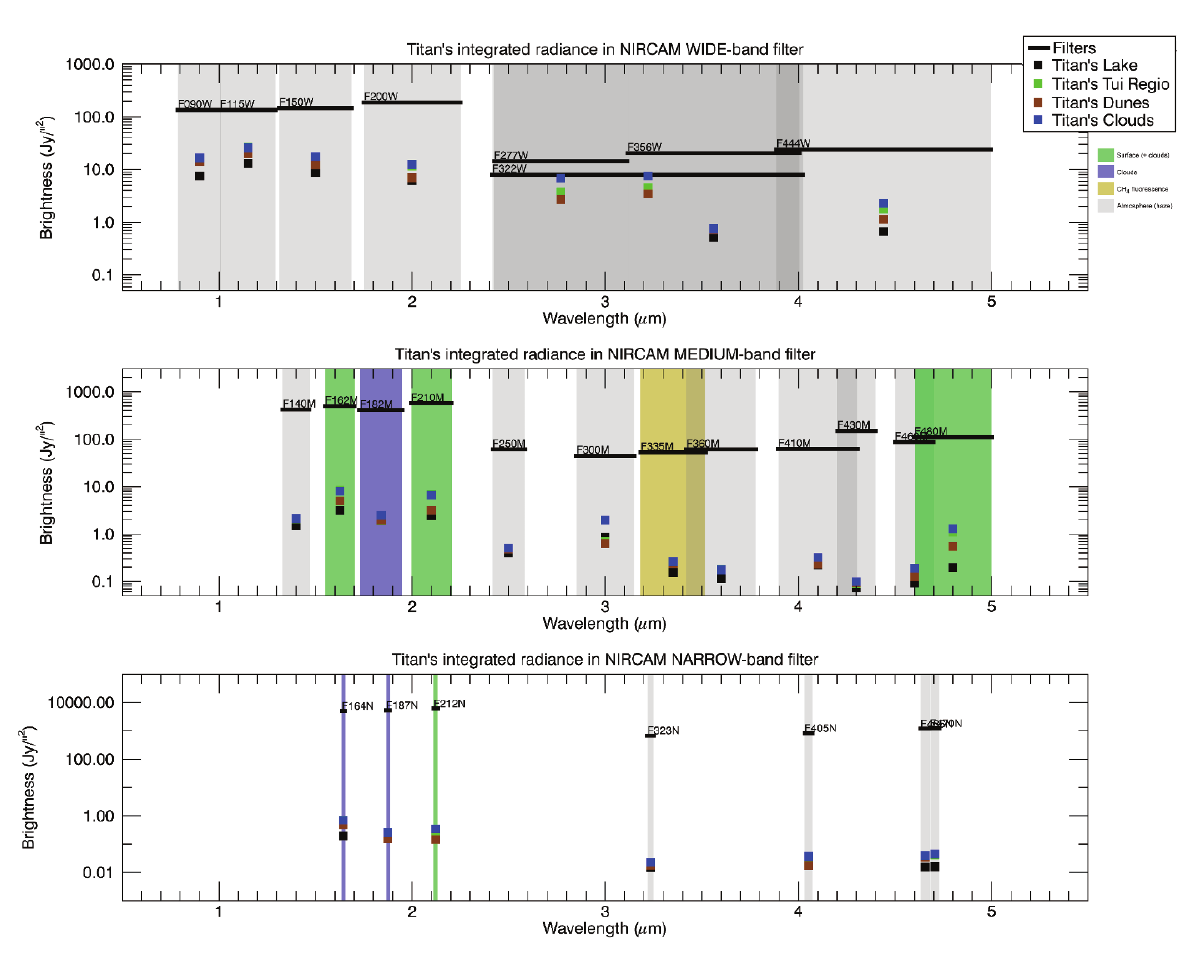}
\caption{Channel-integrated fluxes for various NIRCam filters based on Cassini VIMS spectra, showing how Titan's flux compares to the saturation thresholds in each case. Saturation thresholds have been increased by a factor of 39 from full-frame limits to reflect expected sub-arrayed with $160^2$ pixel window for faster read-out (see text for details).}
\label{fig:nircamsat}
\end{figure}

Fig.~\ref{fig:nircamsat} shows the spectral positioning and expected saturation thresholds of NIRCam filters compared to Titan's near-infrared spectrum (from Cassini VIMS) integrated across each interval. Saturation thresholds have been increased by a factor of 38.8 relative to the full-frame limits, to reflect use of a $160^2$ pixel sub-array. With this smaller sub-array, no bands are saturated - not even the widest filter bands. In contrast, if the sub-arraying is not used, we find that all wide filter bands are saturated, medium band filters are intermediate, and only the narrow-band filters are safely unsaturated. This is an important result for would-be observers to note.

\begin{figure}
 \includegraphics[scale=1.5]{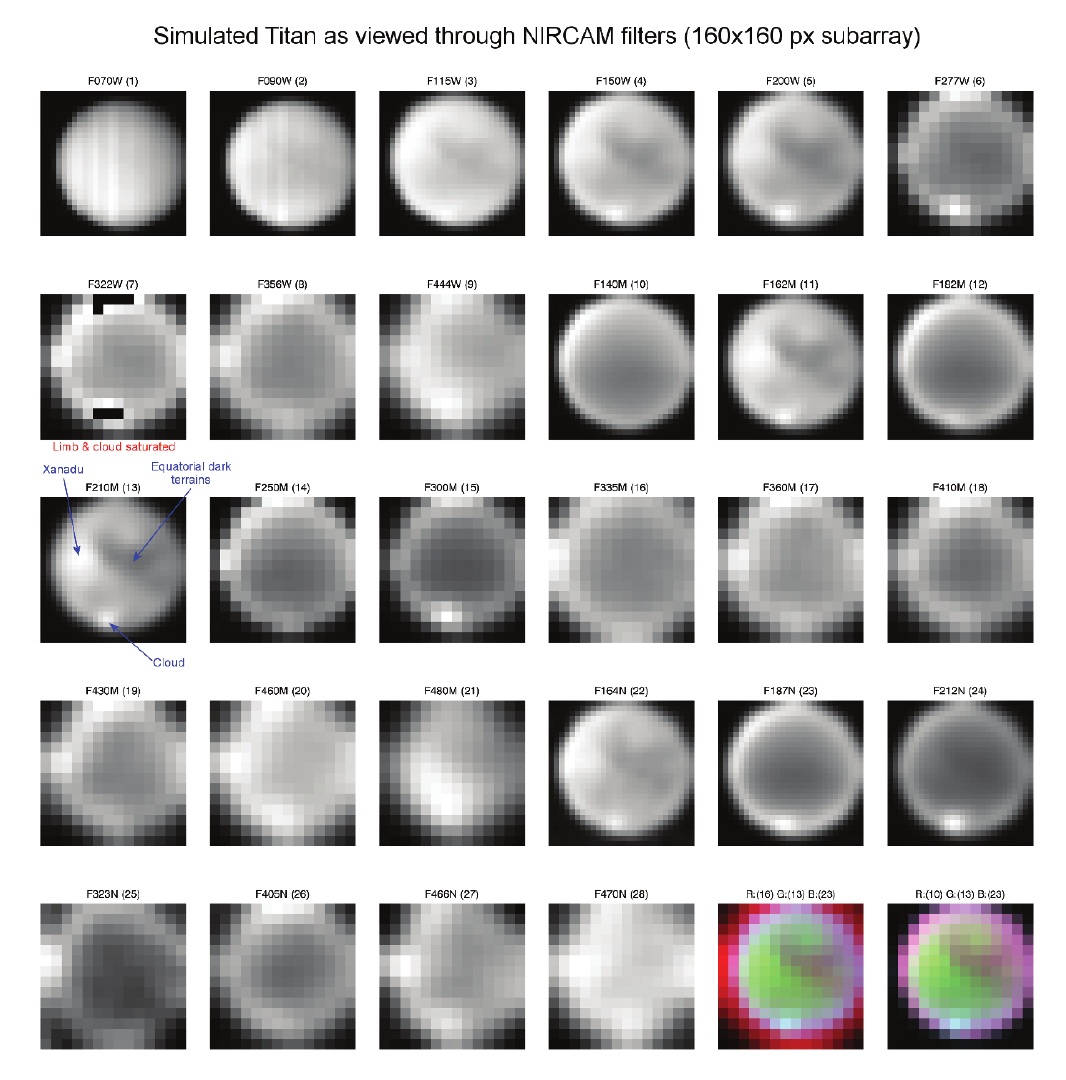}
\caption{Titan VIMS image cube, spectrally integrated across the JWST NIRCam filter set, degraded to diffraction-limited spatial resolution at the filter center. Regions of saturation are indicated, but these are few due to use of $160^2$ pixel sub-array windows. The last two images show RGB color images with the three filters used for each color plane are indicated in parentheses.}
\label{fig:nircamimag}
\end{figure}

Corresponding simulated images are shown in Fig.~\ref{fig:nircamimag}, produced by spectrally integrating a VIMS cube image across the ranges of the NIRCam filters and degrading the spatial resolution to the diffraction limit (1.22$\lambda /D$) at the center wavelength of each window. NIRCam medium-band filters (F210M) and narrow band filters (F164N, F187N, F212N) in the atmospheric windows will be especially helpful to assist with the detection and mapping performed by NIRSpec. Surface imaging and spectroscopy should be attempted after cloud events (as often as feasible) to try to catch surface changes related to rainfall. Physical characterization of clouds and cloud-related surface changes will require radiative transfer analysis. Detection and characterization of high-altitude icy/liquid HCN clouds \citep{dekok14} may also be possible in the near-infrared (3.0--3.1 and 4.8--5.1 \micron\ spectral regions) complemented by gas measurements with MIRI.

Long-term monitoring of seasonal changes in weather patterns will require frequent (weekly - monthly) observations with short integrations ($<$1 min). The model prediction in Fig.~\ref{fig:cloudmodel} extends to 2030 (anticipating an extended lifetime of JWST past the 5-year nominal mission), and shows how JWST could continue the cloud monitoring campaign after the planned end of the Cassini mission around the time of Titan's northern summer solstice in 2017. This continuation of coverage would greatly improve the ability for global climate model-data comparison and deeper insights into the annual and possibly interannual meteorological cycle.

{\bf Figure \ref{fig:nircamsat} here}

{\bf Figure \ref{fig:nircamimag} here}

Table~\ref{tab:clouds} summarizes the scientific goals for Titan clouds that can be achieved on various timescales.

{\bf Table 2 here}

\subsection{Gaseous composition of the troposphere}
\label{sect:troposphere}
{\em Scope: measurements of the spatial/temporal variation of the abundance of methane and measurements of the global abundance of CO and isotopes in the troposphere by near-infrared spectroscopy. Instruments: NIRSpec.}

The primary science aim for tropospheric composition is to measure latitudinal and temporal variations of \methane. In the thermal infrared, Cassini/CIRS \citep{lellouch14} observations have indicated a surprising latitudinal variability of the stratospheric \methane\ mixing ratio, apparently correlated to the latitudes of the most frequent occurrence of tropospheric clouds. 

In the troposphere, the only direct measurement of the methane mixing ratio profile was obtained by the Huygens/GCMS, determining a relative humidity of 45\% at the surface, above which the mixing ratio was constant until saturation was reached \citep{niemann10}. Given the existence of surface temperature gradients with latitude and the occurrence of long-lived dunes at low latitudes and lakes and seas near Titan's poles, the surface methane mixing ratio and humidity can be expected to vary across Titan's surface. In particular, increased insolation during northern summer may be sufficient to produce an increase in humidity due to the evaporation of polar lakes \citep{griffith08}. Summer storms may also be associated with increased (and subsequent decreases in) methane humidity \citep{schaller06b}. 

JWST can be used to search for spatial variations of the methane tropospheric content.
The idea is to use regions of weak \methane\ absorption - either weak bands in the visible or the wings of the methane near-infrared windows - to probe the troposphere. This approach has been previously used by \citet{penteado10a} with the VIMS 0.61~\micron\ band, but results were ambiguous due to the large role of haze opacity at this wavelength. \citet{penteado10b} used the 1.56~\micron\ \dmethane\ band as a proxy for \methane\ abundance and found a constant (to within 20\%) methane abundance below 10~km over 32\dg S--18\dg N. Spatial variations in D/H in tropospheric methane can also be searched for, which could occur due to different saturation vapor pressures for different mass isotopologues.

\begin{figure}
\includegraphics[scale=0.65]{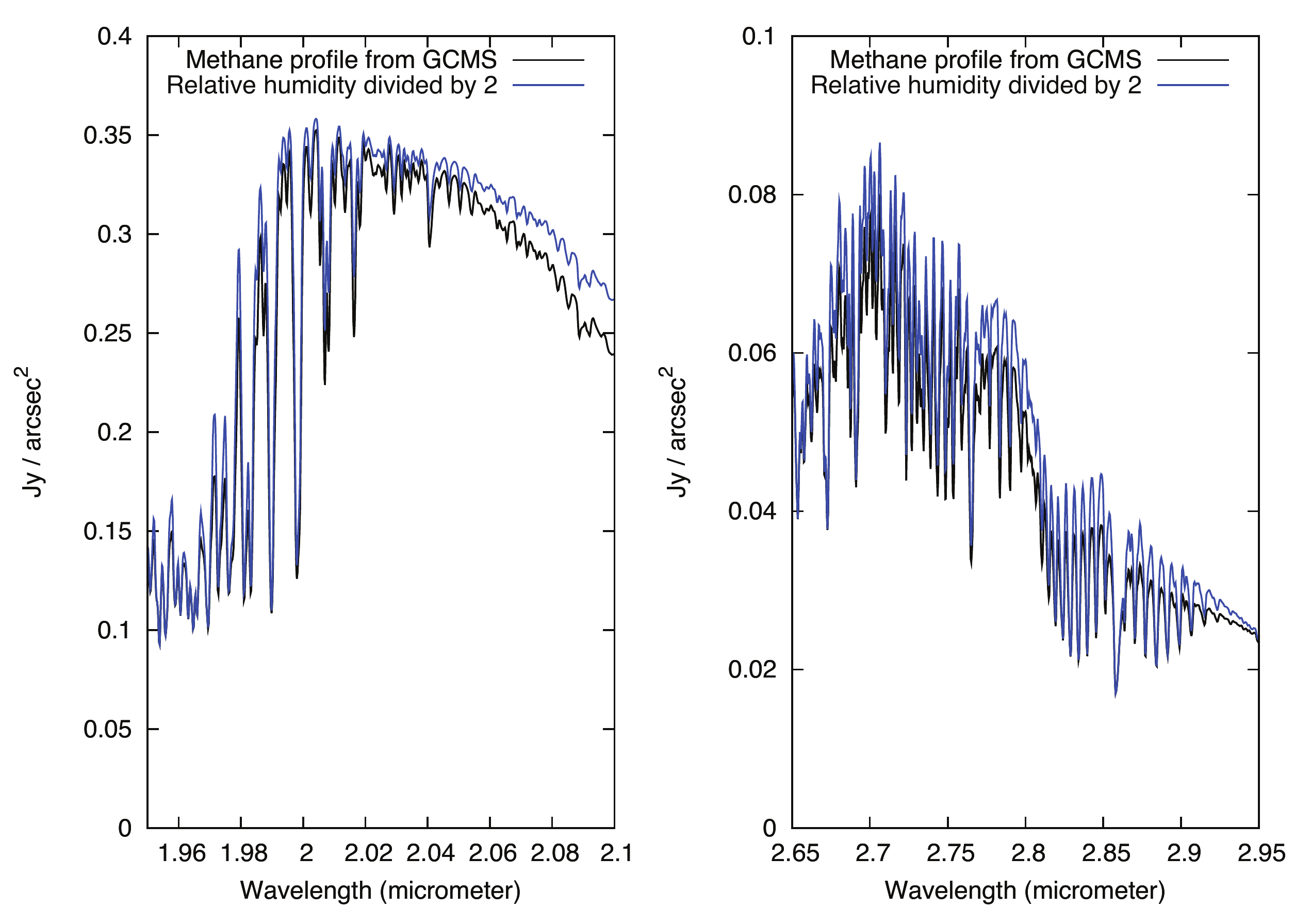}
\caption{NIRspec spectrum computed for maximum spectral ($R=2700$) resolution. The two calculations show the spectrum for both a nominal (Huygens GCMS) and half nominal methane relative humidity profile.}
\label{fig:methanehumid}
\end{figure}

Synthetic spectra calculated for the 1.4--3.2~\micron\ range and at the maximum $R = 2700$ resolving power afforded by NIRSpec, indicate that the wings of the near-infrared methane windows are transparent enough to permit determination of the methane tropospheric column. This is illustrated in Fig.~\ref{fig:methanehumid} in two spectral intervals at 1.95--2.10~\micron\ and from 2.65--2.95~\micron . Spectra have been calculated using both the nominal Huygens GCMS methane abundance profile, and with the relative humidity halved. Interestingly, in general the lines appear to be deeper for the lower \methane\ abundances. This is because the \methane\ tropospheric opacity occurs mostly through far wings of strong lines, which depresses the local continuum. Note that saturation is not expected to be a problem in these intervals at high spectral resolution - see Fig.~\ref{fig:nirspecsat}.
 
{\bf Figure \ref{fig:methanehumid} here}

Additional windows suitable for this task occur at 1.48--1.62 and 1.87--1.90 \micron . Orienting the $0.2\times 3.3${\arcsec} slit of NIRSpec parallel to the polar axis (which does introduce an overhead associated with spacecraft positioning) would permit a search for latitudinal variations of the methane abundance, and a 2-D map could be built by repeating the measurement at a variety of central meridian longitudes. Albeit perhaps less likely, diurnal variations of the methane content could be searched for by orienting the slit parallel to the equator (again, as permitted by the spacecraft). Any methane variations could then be correlated with surface and troposphere temperature measurements e.g. in the far-IR, as well as cloud detections at various wavelengths. Measurements should be performed at a variety of epochs to search for seasonal evolution.

Measurement of the abundance of CO and its isotopes should be possible as well, e.g. using near-IR bands at 1.6 and 4.7~\micron . The far-IR rotational lines measured by Cassini CIRS at $\sim$300~{\micron} will not fall into the JWST range.

\subsection{Gaseous composition of Titan's stratosphere}
{\em Scope: abundances of trace gases in the stratosphere by near-infrared spectroscopy of fluorescent emissions (NIRSpec) and mid-infrared spectroscopy of thermal emissions (MIRI). Spatial and temporal monitoring, and search for new gases and isotopes. Instruments: NIRSpec, MIRI.}
\label{sect:stratosphere}

\subsubsection{Mid-Infrared Observations}

{\bf Figure \ref{fig:cirsspec} here}

Cassini CIRS confirmed and extended the Voyager discoveries about Titan's atmospheric composition, showing that Titan has an extremely rich stratospheric chemistry, with emission from many hydrocarbon and nitrile species (Fig.~\ref{fig:cirsspec}) \citep{coustenis07, coustenis10}. JWST's MIRI instrument will expand spectral capability relative to Cassini CIRS, although more restricted in other respects such as spatial resolution, polar coverage, and capability for limb sounding. Differences include the wavelength coverage, extending the spectral range down to 5~\micron\ (CIRS stops at 7~\micron ); and a spectral resolution that somewhat exceeds that of CIRS especially towards long wavelengths. E.g. at 7~\micron\ the $R\sim 3250$ resolving power of MIRI MRS yields a resolution of 0.4~\cm\ (cf. CIRS 0.5 \cm ), while at 28~\micron\ the MRS resolving power of 1550 achieves a resolution of 0.25~\cm\ (cf. CIRS again 0.5~\cm ). The higher resolution may yield improved detections of weak gas bands and better separation of gas bands that are mixed together at CIRS resolution \citep[e.g. propane and propene at around 11~\micron ,][]{nixon13}.

Fig.~\ref{fig:miri} shows a synthetic Titan spectrum compared to MIRI MRS saturation and sensitivity levels. Titan provides an extremely bright target for JWST and will saturate the detectors in the 12--16~\micron\ and 24--28~\micron\ regions. The former is a rich emission region for many of Titan's more abundant hydrocarbons and nitriles, especially stratospheric \ethane , \acet , \methacet , \diacet\ and HCN, with minor contributions from \cyanoacet , \coo , and \benzene\ \citep{teanby07, teanby08a, vinatier07a, vinatier07b, vinatier10a, vinatier10b}. These species will be challenging to observe with JWST, requiring reduced integration times (sub-array mode with the MRS if permitted) to avoid saturation. An unsaturated region around 15~\micron\ is promising for HCN, \acet , and \methacet . The 5--12~\micron\ region is highly suitable for measuring composition and temperature as this region maintains a high S/N without saturating the detectors. Emission from \methane , \ethylene , \propane\ and \ethane\ \citep{nixon09b} , along with minor species and isotopic lines such as \dmethane\ \citep{bezard07, nixon12b} are viable science targets. This region also has the highest spatial resolution with a pixel scale of 0.2--0.3{\arcsec}. It thus provides the opportunity for limited spatial temperature and composition mapping.

{\bf Fig.~\ref{fig:miri} here}
 
In summary, high-level MIRI science goals include:

\begin{enumerate}
\item {\em Measurement/detections of trace species:} The high S/N in the 5--12~\micron\ range makes detection of new species and minor isotopes a possibility \citep{nixon10b}, even with modest integration times. Limited spatial mapping of known hydrocarbons and nitriles should also be possible. 
\item {\em Temporal variations in temperature:} MIRI will be able to infer stratospheric temperatures through measurements of the $\nu_4 $ band of \methane\ at 7.7~\micron , and assuming the abundance of \methane\ \citep{achterberg08a}. Although spatial resolution will be low (3--4 samples across the disk), it will be sufficient to track changing temperatures in each hemisphere, extending the seasonal dataset of CIRS \citep{achterberg11} to the northern summer and fall. 
\item {\em Temporal variations in composition:} MIRI will be able to track large scale variations in composition throughout the mission \citep{teanby09b, teanby10a, teanby12, vinatier15}, separating northern and southern hemispheres at the shorter wavelengths (5--12~\micron , see Fig.~\ref{fig:resolution}). This will provide important constraints on photochemical/dynamical models. Some limited work around 15~\micron\ may also be possible, but with lower spatial resolution.
\end{enumerate}

\subsubsection{Titan's middle atmosphere in the near-infrared}
Near-infrared measurements of Titan's dayside emission by fluorescence provide another important source of gas abundance information.
HCN has already been observed from ground-based telescopes near 3.0~\micron\ at high spectral resolution ($R\sim 20000$) \citep{geballe03, yelle03, kim05}.  From that spectral region they probe the HCN concentration near the upper stratosphere (200 to 400~km). Similarly, JWST should be capable of measuring the HCN concentration at those levels when observing Titan in the daylight (see Fig.~\ref{fig:nadir}, top left panel), complementing abundance measurements in the mid-IR by MIRI. Similar arguments apply to \acet , whose concentration was also derived from ground-based observations in this spectral region \citep{kim05}. As shown in the top-left panel of Fig.~\ref{fig:nadir}, \acet\ abundances could be derived from JWST observations of an illuminated Titan for a reasonable integration time. Latitudinal/seasonal variation of those species in the upper stratosphere/lower mesosphere as shown by \cite{vinatier15} could be achievable with JWST.

 {\bf Figure \ref{fig:nadir} here}

Methane emissions in the 3.3~\micron\ region at high spectral resolution ($R\sim 20000$) \citep{kim00, kim05}
have also been observed using ground-based telescopes. The first was significantly contaminated by the telluric contribution. Nevertheless, from the observation near 3.3~\micron\ they derive the temperature of upper stratosphere/ lower mesosphere (500--600~km). 
Fig.~\ref{fig:nadir} (top right panel) shows nadir simulations where the fundamental and first hot bands can be measured with a high S/N ratio. The telluric-free observations of \methane\ will then enable better measurements of the middle atmosphere temperatures. Also, the non-LTE contributions of the different \methane\ bands will be better constrained. 
The weaker methane emission near 3.6--3.9~\micron\ (Fig.~\ref{fig:nadir}, bottom left panel) could also be detectable by JWST. For this band there is little day/night difference and it could be observable at both times. This weaker band could provide information about temperature at lower altitudes than the 3.3~\micron\ bands.

{\bf Figure \ref{fig:vimslimb1} here}

Early measurements of the CO abundance in Titan's stratosphere ranged over a factor two in values \citep{hidayat98, lellouch03, gurwell04, lopez-valverde05, bailey11}. 
Although Cassini CIRS measurements in the lower stratosphere have reduced uncertainties down to $\pm$20\% \citep{dekok07a}, the highest possible precision is required because CO is expected to be horizontally uniform due to its extremely long chemical lifetime, and therefore may be used as an additional thermometer in Titan's atmosphere.
Observations of the CO 4.7~\micron\ emission at high spectral resolution have already been made using ground-based telescopes \citep{lellouch03, lopez-valverde05}, proving that stratospheric CO can be inferred. CO has also been derived by using VIMS nadir night-time spectra at 4.7~\micron\ but with a lower spectral resolution ($R\sim 300$) \citep{baines06}, and it is also being currently retrieved from VIMS daytime limb spectra (Fig.~\ref{fig:vimslimb1}). JWST nadir observations at $R\sim 2700$ in the 4.7~\micron\ region (see Fig.~\ref{fig:nadir}, bottom right panel) could therefore be very useful for a re-determination of the CO stratospheric abundance.
 
{\bf Figure \ref{fig:vimslimb2} here}
 
Typical daytime limb spectra taken by Cassini/VIMS at several tangent heights are shown in Fig.~\ref{fig:vimslimb2}. VIMS daytime limb spectra show very prominent features of HCN and \acet\ near 3.0~\micron , the strong 3.3~\micron\ \methane\ (fundamental and hot) bands, \methane\ bands from the dyad levels near 3.9~\micron , CO emission near 4.7~\micron\ and of \dmethane\ near 4.5~\micron . Vertical profiles of HCN and \methane\ concentrations in Titan's upper atmosphere have already been derived from these spectra \citep{adriani11, garcia-comas11, maltagliati15}. Similar inversions for \acet\ and CO and \dmethane\ are in progress. In addition, large concentrations of PAHs have been found in Titan's upper atmosphere \citep{dinelli13, lopez-puertas13} - a key finding for understanding the formation of aerosols, which likely begin forming in the upper atmosphere where ion chemistry plays a key role \citep{lavvas13}. 
 
Although JWST will not be able to resolve Titan's limb nearly as well as Cassini VIMS, it may be possible to separate the lower and upper atmospheric limb for bright fluorescent emissions. Using a notional resolution of 0.1{\arcsec} for NIRSpec IFU mode corresponds to a spatial pixel size of $\sim$650~km, shown in Fig.~\ref{fig:nirspecIFUresol}, and sufficient for this purpose. Fig.~\ref{fig:nirspec_sensi} shows the computed sensitivity for NIRSpec. For an exposure time of $10^5$~s and S/N of 10, the minimum detectable flux is $\sim$1000 nJy, or $1.0^{-4}$~Jy/arcsec$^2$ for 0.1{\arcsec} pixels. This demonstrates that the radiances of Figs.~\ref{fig:vimslimb1} and \ref{fig:vimslimb2} would be observable. Long integration times on the limb might cause saturation in surface pixels however.

{\bf Figure \ref{fig:nirspec_sensi} here}

In addition, JWST NIRSpec can be operated with a much better spectral resolution than VIMS, and so will provide unique information. E.g. IFU mode with NIRSpec provides spectral resolution up to 2700 in the 2.9--5.0~\micron\ (filter F290LP) spectral range. This will enable HCN and \acet\ features near 3.0~\micron\ and the 3.3~\micron\ \methane\ bands to be better resolved (compare Fig.~\ref{fig:nadir} with Figs.~\ref{fig:vimslimb1} and \ref{fig:vimslimb2}), potentially providing crucial information about which PAHs are really present in Titan upper atmosphere. These compounds emit around 3.28~\micron, which is coincident with the R-branch of the strong \methane\ $\nu_3$ fundamental band.

\subsection{Middle-atmosphere clouds and hazes}
\label{sect:hazes}
{\em Scope: multi-spectral monitoring of Titan at opaque near-infrared wavelengths to determine spatial and temporal variations in haze distribution, and to make inferences about aerosol composition and atmospheric dynamics. Instruments: NIRSpec, NIRCam.}

JWST observations of Titan's haze would build on the legacy of observations from ground-based, HST, Pioneer, Voyager and Cassini platforms. Cassini especially has provided a wealth of imaging and spectral data free from telluric absorption. Of most interest for study of Titan's haze using JWST will be high resolution ($R\sim 2700$) spatially resolved spectra using the IFU mode with NIRSpec. With IFU, Titan's disk can be sampled by multiple apertures with angular scale of 0.1{\arcsec}. At such angular resolutions it will be possible to isolate hemispheric differences, Titan's polar hood region, and occasional large regional cloud formations (Fig.~\ref{fig:arrowcloud}). The spectral resolution and range available will provide roughly 100~km vertical resolution. The main haze extends from the surface to $\sim$1000~km altitude, but at near-infrared wavelengths only the lowest 500~km is optically dense enough to sense. That range is interesting because it encompasses the troposphere to upper stratosphere, where the radiative time constant varies from about 30 years near the surface to about four months near 500~km. The exact mechanisms for how seasonal change occurs at these vastly different scales and altitudes and manifests in the form of haze profiles as a function of latitude and time remains obscure.

{\bf Figure \ref{fig:haze} here}

There is a local minimum in the vertical profile of the haze, which produces the appearance of a seasonally varying `detached' haze layer seen by both Voyager and Cassini (Fig.~\ref{fig:haze}).  Cassini observations showed that the detached layer dropped in altitude over time from about 500 km in 2007 to about 380 km in 2010, more consistent with the altitude of $\sim$350~km observed by Voyager 1 and 2 in 1980--1981 \citep{west11}. The drop was most rapid near equinox, and, at the time of writing, the detached haze is not observed. Unfortunately the spatial scale of this feature is too small for JWST to resolve. However, other important seasonal phenomena, such as hemispherical changes in haze distribution \citep{lorenz04, sromovsky81} and formation and evolution of a polar ethane cloud encompassing as much as 60\dg --90\dg N \citep[][]{griffith06b}, would be observable by JWST. Cassini observations also reveal other seasonally related phenomena such as a cloud patch that formed in the southern (winter) polar region after equinox \citep[Fig.~\ref{fig:spolecloud},][]{west15}, suggested to be composed of condensed HCN ice/liquid \citep[][]{dekok14}. Because these regions are in shadow and tilted away from Earth JWST might not be able to observe similar phenomena after the next equinox period. However, Cassini observations show that the wind field is tilted by about 4\dg\ with respect to the solid body spin axis \citep{achterberg08b}, providing better visibility as seen from Earth for features that have some azimuthal asymmetry such as the polar cloud mentioned above.

{\bf Figure \ref{fig:spolecloud} here}

In the thermal infrared, Cassini observations probe haze opacity in the middle atmosphere \citep[100--300 km altitude;][]{vinatier10b}. Although the MIRI instrument will only be able to resolve hemispheric differences, such observations can also be important for studying the seasonal behavior of the haze. Note that JWST has continuous spectral coverage through the infrared with NIRSpec and MIRI, whereas Cassini has a spectral gap at 5--7~\micron\ between the VIMS and CIRS spectral ranges \citep[c.f.][Fig. 5]{vinatier12}. JWST will therefore provide crucial first-of-a-kind information on the gases and hazes in this region.

\section{Summary and Conclusions}
\label{sect:conc}

In this report, we have described the potential of the JWST to carry out scientific investigation of Titan, focusing on the capabilities of NIRSpec, NIRCam, and MIRI. The abilities of the fourth instrument NIRISS, which are highly complementary to NIRCam and NIRSpec, will undoubtedly expand the scientific potential still further as unique uses are found. Here we summarize our findings on the scientific topics, and address the question of the place of JWST in the wider context of Titan remote sensing investigation. We conclude with some technical comments addressed to the JWST implementation team that we hope will further improve the already strong possibilities for Titan science with JWST.

\subsection{Summary of key findings}
JWST has excellent capability to conduct scientific investigation of Titan, including the following five core scientific areas:

\begin{description}
\item[Surface:] JWST can spectrally characterize Titan's surface especially with NIRSpec, providing new discriminants of surface solid and liquid composition, and monitoring changes in albedo that are due to rainfall. NIRCam can be used to provide context images for the spectra acquired by NIRSpec. At these instruments' spatial resolution, several pixels of one observation would result in the mixing of different compositional units. The unmixing of the pixels could be made by cross-comparing JWST images with Cassini VIMS or ISS higher resolution images to determine surface percentages of different units in a single pixel.
\item[Clouds:] JWST can monitor the apparition and frequency of lower atmosphere clouds, using NIRSpec and NIRCam, providing information not only on coverage and latitudinal and temporal distribution necessary to constrain climate models, but also potentially on cloud altitude and chemical composition (Table~\ref{tab:clouds}).
\item[Lower atmospheric composition:] JWST will search for spatial and temporal variation in the relative humidity of methane, giving important insights in meteorological and/or chemical sources and sinks of this key gas species.
\item[Middle atmosphere composition:] JWST can extend the science of Cassini CIRS and VIMS. In the mid-infrared, MIRI will continue the legacy of CIRS by searching for new gas species using higher spectral resolution and monitoring changes in disk-averaged gas abundances over several years. In the near-infrared, NIRSpec will be able to measure dayside gas fluorescence of HCN, CO, \acet , \methane\ and other species, complementing the findings of VIMS and ground-based telescopes.
\item[Middle atmosphere hazes and clouds:] JWST can monitor the large-scale distribution of Titan's haze, as it switches in appearance between hemispheres in response to annual circulation and insolation. 
\end{description}

\subsection{JWST in the context of Titan remote sensing}

With the demise of the Cassini spacecraft in September 2017, a major gap will open up in the seasonal monitoring of Titan. Cassini by then will have orbited Saturn for 13.25 years, almost half a Titanian year (14.25 Earth years), and made 127 targeted flybys of Titan during this time at ranges typically of $\sim$1000~km, well within the ionosphere. Despite not being a Titan orbiter, Cassini will have imaged the vast majority of Titan's surface ($>$90\%) at a resolution of 5~km or better, with SAR RADAR coverage of $\sim$40\% at a resolution of several 100~m \citep{lorenz13}. Cassini's twelve on-board instruments have examined Titan in unprecedented detail, from the distant magnetospheric interactions with the Saturnian field and solar wind, to the interior, via gravity measurements. In between the ionospheric composition has been directly sampled, measuring ions and neutral particles, the upper neutral atmosphere has been probed with UV occultations and IR fluorescence, the middle and lower atmosphere have been investigated with thermal IR spectroscopy and radio occultations, and the surface has been mapped at multiple short and long wavelengths. Clearly, the capability for high spatial, temporal and phase angle coverage of Titan that will be lost with the end of the Cassini mission will be considerable.

In this context, JWST will have a significant impact - not a replacement for Cassini, but nevertheless a highly capable facility for monitoring Titan during its southern winter season, during the JWST mission 2018-2028. These capabilities have been described in detail in earlier sections of this paper. In general, JWST will restore temporal monitoring, and have enhanced spectral resolution at near-IR wavelengths relative to Cassini. The main losses include spatial resolution, plus the ability for in-situ sampling and occultations (at multiple wavelengths), and active science such as RADAR and radio bi-static experiments.

Other large ground and space-based facilities will provide complementary capabilities. HST, VLT, Keck and other large optical telescopes can be turned to observe cloud outbreaks \citep{brown02, gibbard04, roe05, schaller06a, schaller06b}, as detected by smaller monitoring telescopes. Mid- and far-infrared spectroscopy will be possible with SOFIA, the airborne observatory. Although spatially unresolved, instrumentation such as EXES, with resolving power of $R\sim10^5$, will offer the possibility of separation and detection of new gas species from stronger overlying infrared bands. Perhaps the most exciting prospect for long-term monitoring of Titan's atmospheric gases will come from ALMA, the recently operational sub-millimeter observatory, which has proven capability to map Titan in the sub-millimeter using baselines of 1.5~km or less \citep{cordiner14, cordiner15}. In the near future, expanded baselines to 15~km will permit spatial resolutions on Titan of 100~km, enabling high fidelity monitoring of seasonal changes in some gas species (those that have dipoles, and therefore rotational spectra).

\subsection{Technical comments}
Several items for consideration are noted. 

The first is the inability of JWST to observe Titan at every desired epoch due to solar elongation restrictions. This restriction cannot be overcome as it is a fundamental limit of the observatory design. In this case, it is hoped that proposals to observe Titan during visibility windows will be given due priority, since there will be long periods when Titan cannot be observed.

The second concern is the saturation of all three instruments (NIRCam, NIRSpec, MIRI) in certain high-flux regions of Titan's spectrum at standard read-out times. 
NIRCam has the ability to overcome this limitation by use of (a) narrow spectral filters and standard read-out times, although this reduces available spectral ranges, or more generally (b) sub-arraying to avoid saturation, which should work for all filters. We find that a sub-array of 160$^2$ pixel is ideal. NIRSpec can be used at resolutions of $\geq1000$ with the IFU mode to avoid saturation. Saturation in certain spectral regions with MIRI may be more problematic to overcome. We do note however that all saturation limits are based on specification and initial characterization of the instruments.  After launch, real characterization will begin, assessing scattered light levels and real instrument performance (expected in time for the second call for proposals). Software updates and new operational modes may also be implemented that will provide more versatile options than considered in this paper.

We further suggest that observing proposals be permitted for `picket-fence' campaigns of atmospheric and surface monitoring, that would allow for frequent, short-duration visits to Titan (several minutes with NIRCam and a few seconds-minutes with NIRSpec) to search for sporadic clouds or surface changes. In the event of clouds being detected, observers could follow up with TOO requests. Short-duration observations using the NIRSpec instrument would also provide critical improvements to better constrain the composition of Titan's surface than has been possible with Cassini.

\acknowledgments
{\large \bf Acknowledgements}

\begin{figure}
\includegraphics[scale=0.65]{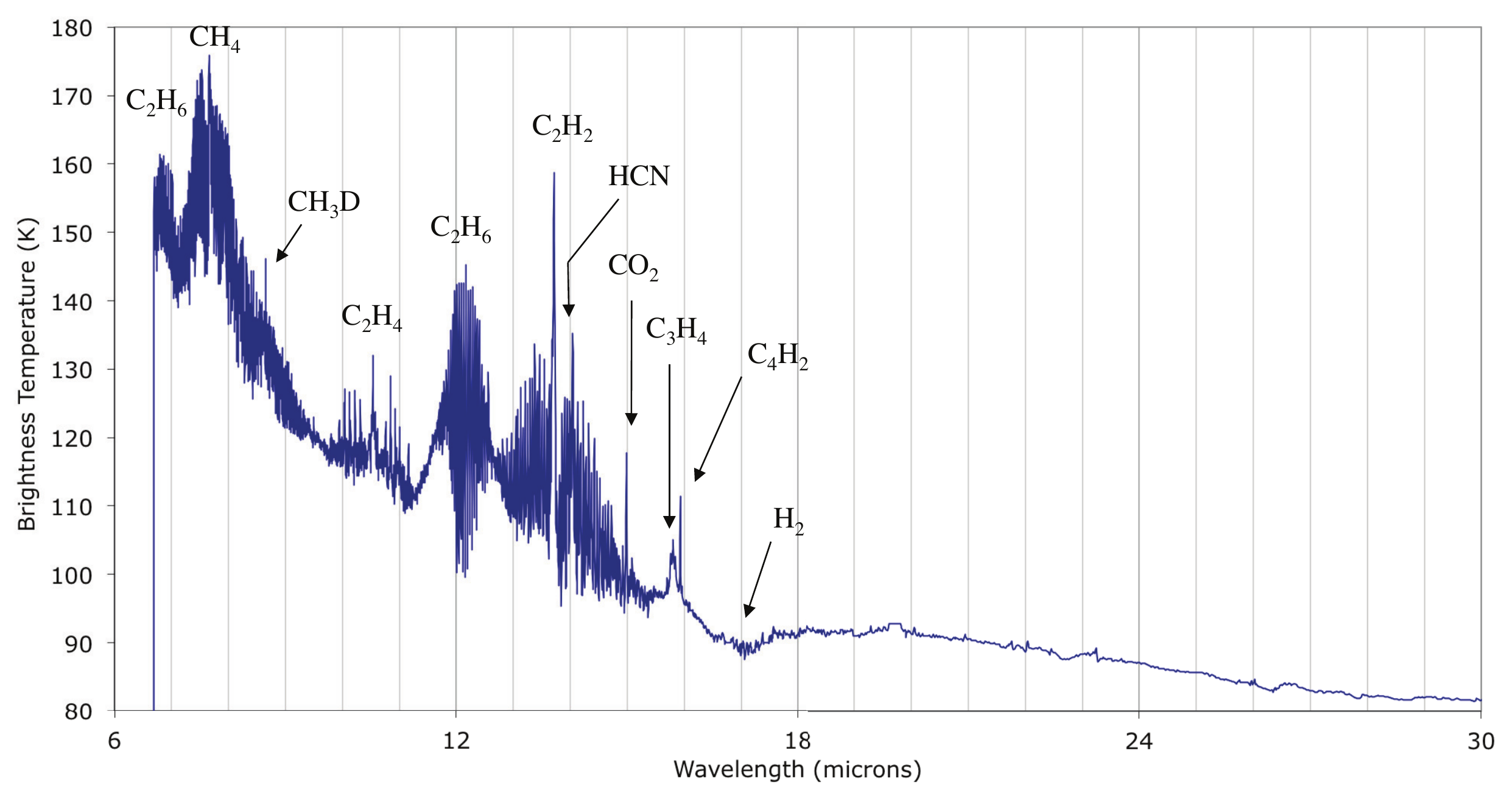}
\caption{Titan infrared spectrum from Cassini CIRS at low latitudes, cropped to cover the MIRI spectral range. Locations of notable gas emission bands are noted. At winter polar latitudes, the spectrum is enhanced with short-lived species and several new emissions emerge, such as \benzene\ at 15~\micron .}
\label{fig:cirsspec}
\end{figure}

\begin{figure}
\includegraphics[scale=0.33]{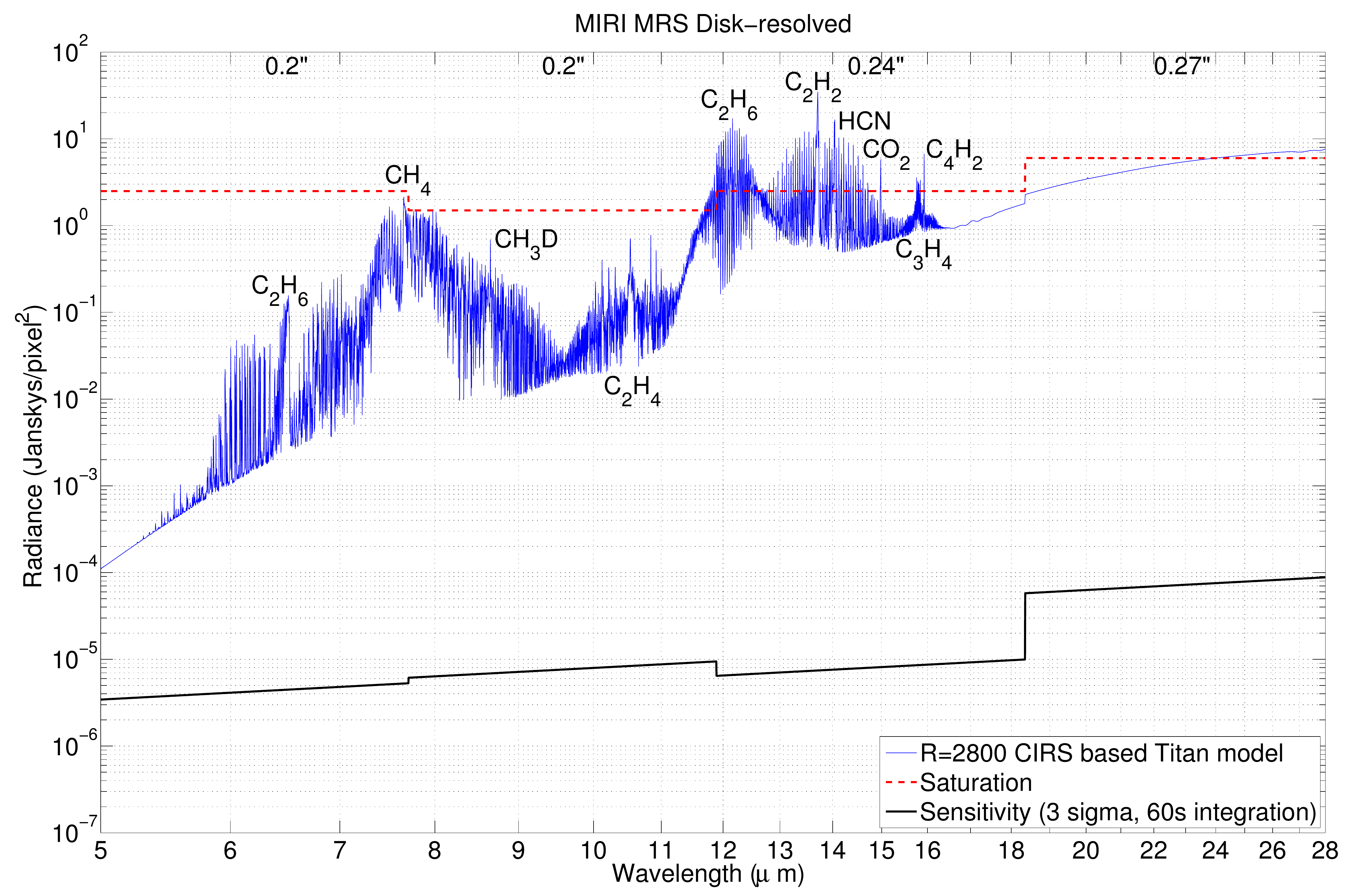}
\caption{Synthetic spectrum of Titan generated using an atmospheric model based on Cassini CIRS data. S/N is very favorable except for the 12--16~\micron\ and 24--28~\micron\ regions, which saturate the detectors. Numerical labels at the top give assumed pixel sizes of the four MIRI sub-arrays.}
\label{fig:miri}
\end{figure}

\begin{figure}[ht]
\noindent
\hspace{-2.2cm}
\begin{tabular}{ll}
\rotatebox{-270}{\includegraphics[width=70mm]{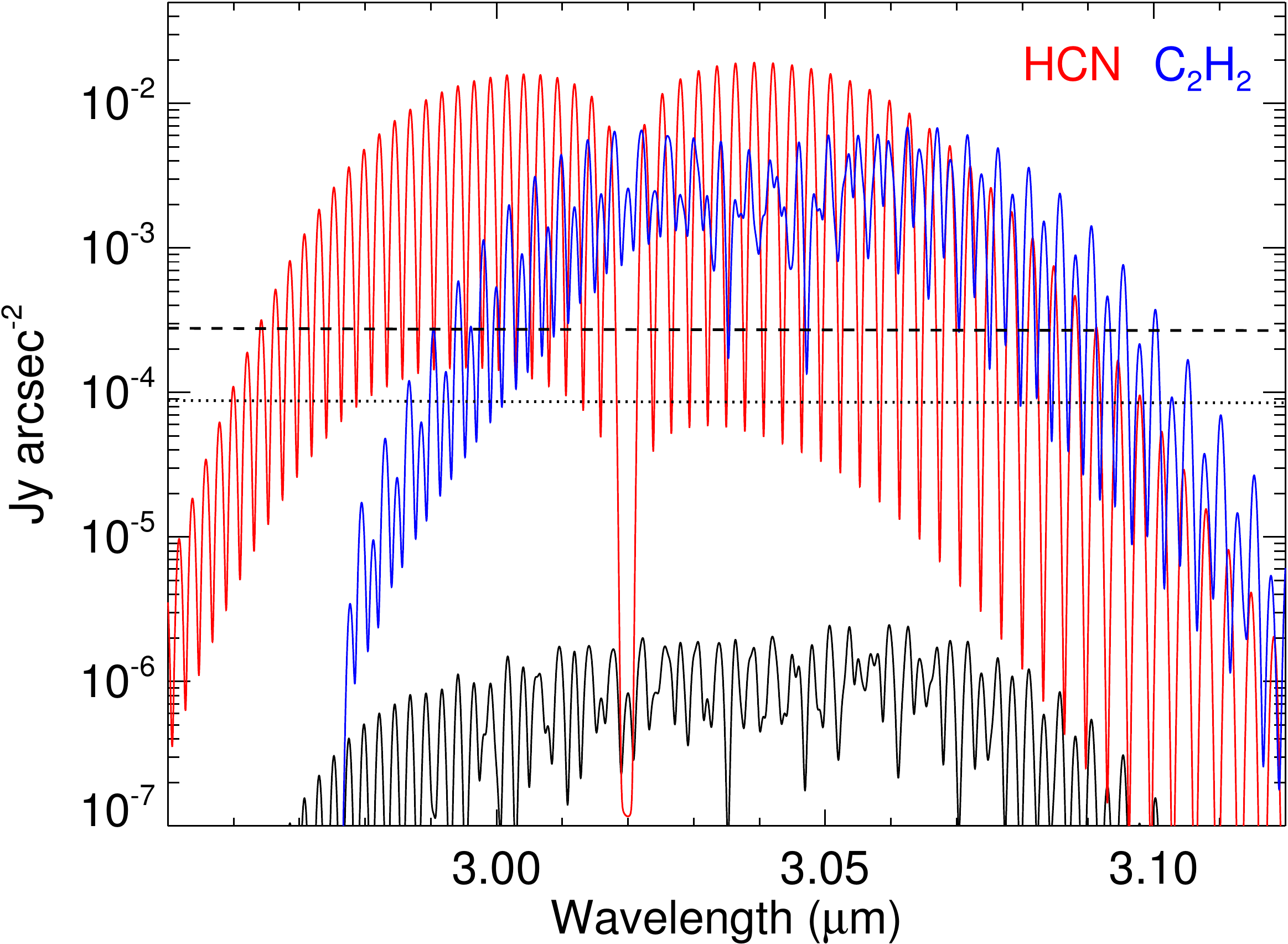} } &
\rotatebox{-270}{\includegraphics[width=70mm]{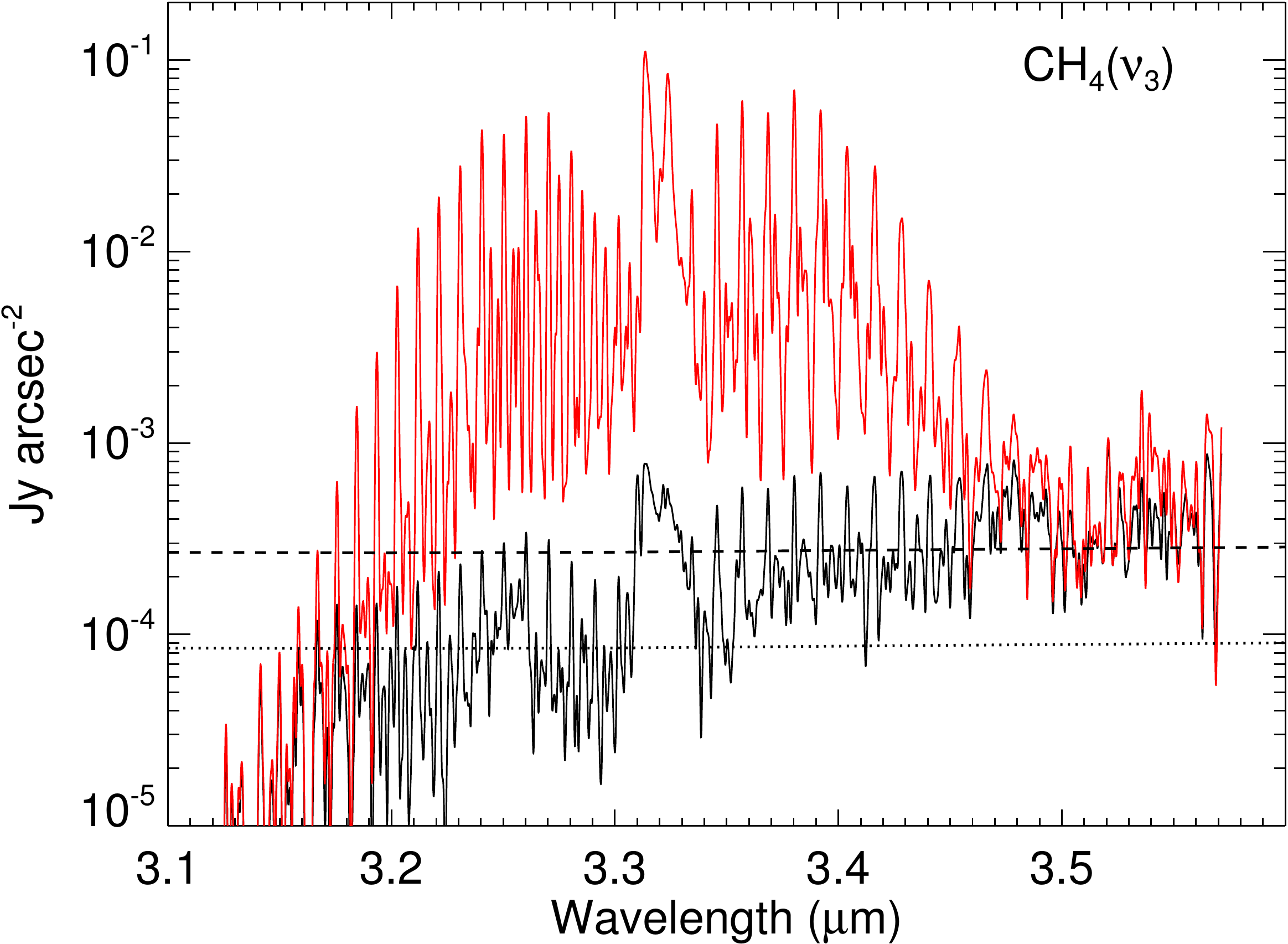} } \\
\rotatebox{-270}{\includegraphics[width=70mm]{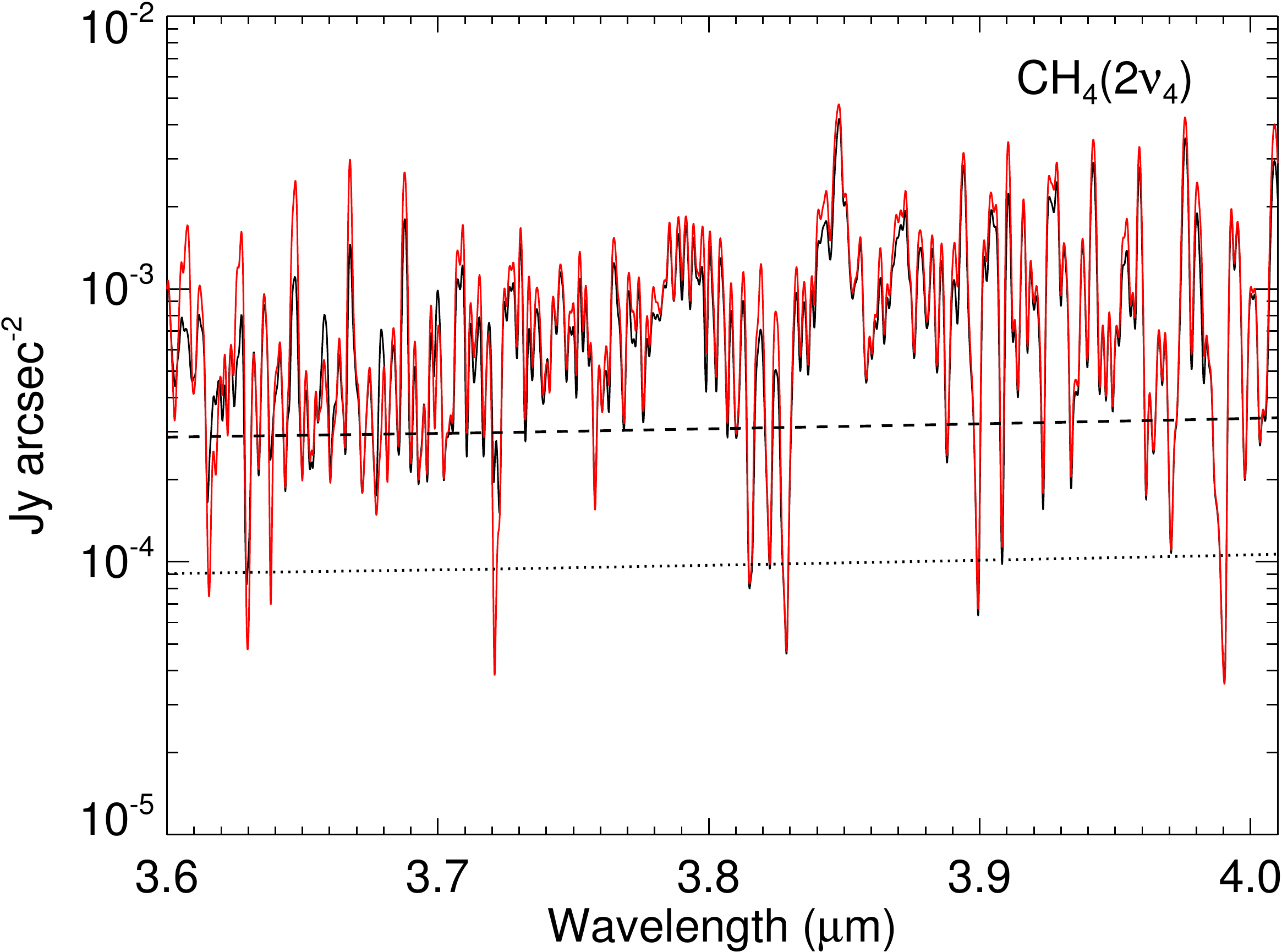} } & 
\rotatebox{-270}{\includegraphics[width=70mm]{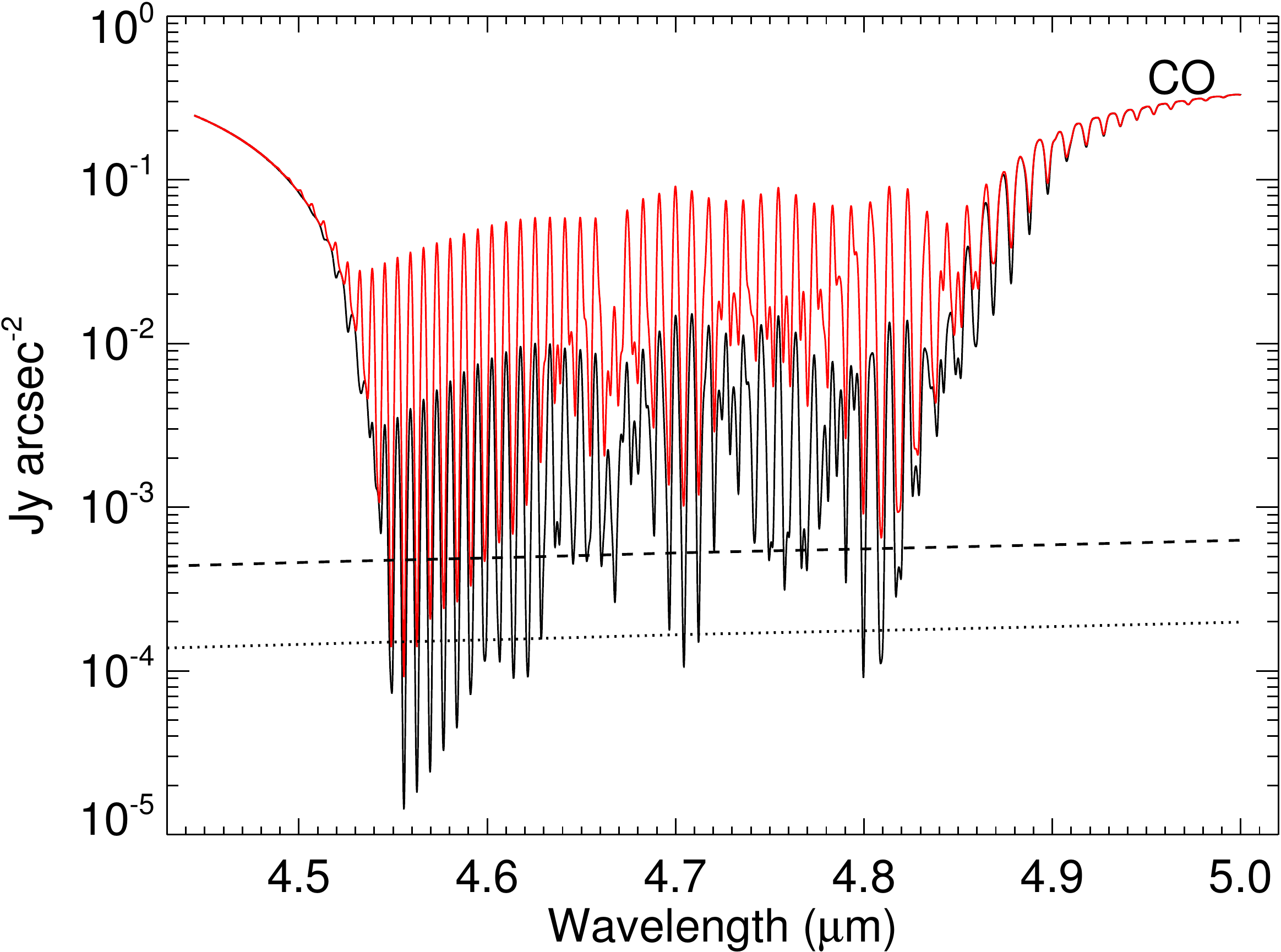} } 
\end{tabular}
\caption{Titan nadir radiance simulations at a resolving power $\sim$ R=2700 for several spectral regions. Except for the HCN and \acet\ case (top left), the others include, in addition to the atmospheric contribution, the reflected solar component. The radiances incorporate non-LTE (non-local thermodynamic equilibrium) populations of the corresponding emitting levels, computed for daytime conditions as described by \citet{adriani11} for HCN, \citet{garcia-comas11} for \methane , and unpublished results for \acet\ and CO. Daytime non-LTE radiances are in red and blue for \acet\ and the LTE radiance (roughly corresponding to nighttime conditions) in black. The dotted and dashed lines show the NIRSpec sensitivity for the spatial resolution of $0.1\times 0.1$ arcsec$^2$/pixel, a S/N=10, and exposure times of 10$^5$\,s and 10$^4$\,s, respectively.}
\label{fig:nadir}
\end{figure}

\begin{figure}[ht]
\begin{center}
\rotatebox{-270}{\includegraphics[width=11cm]{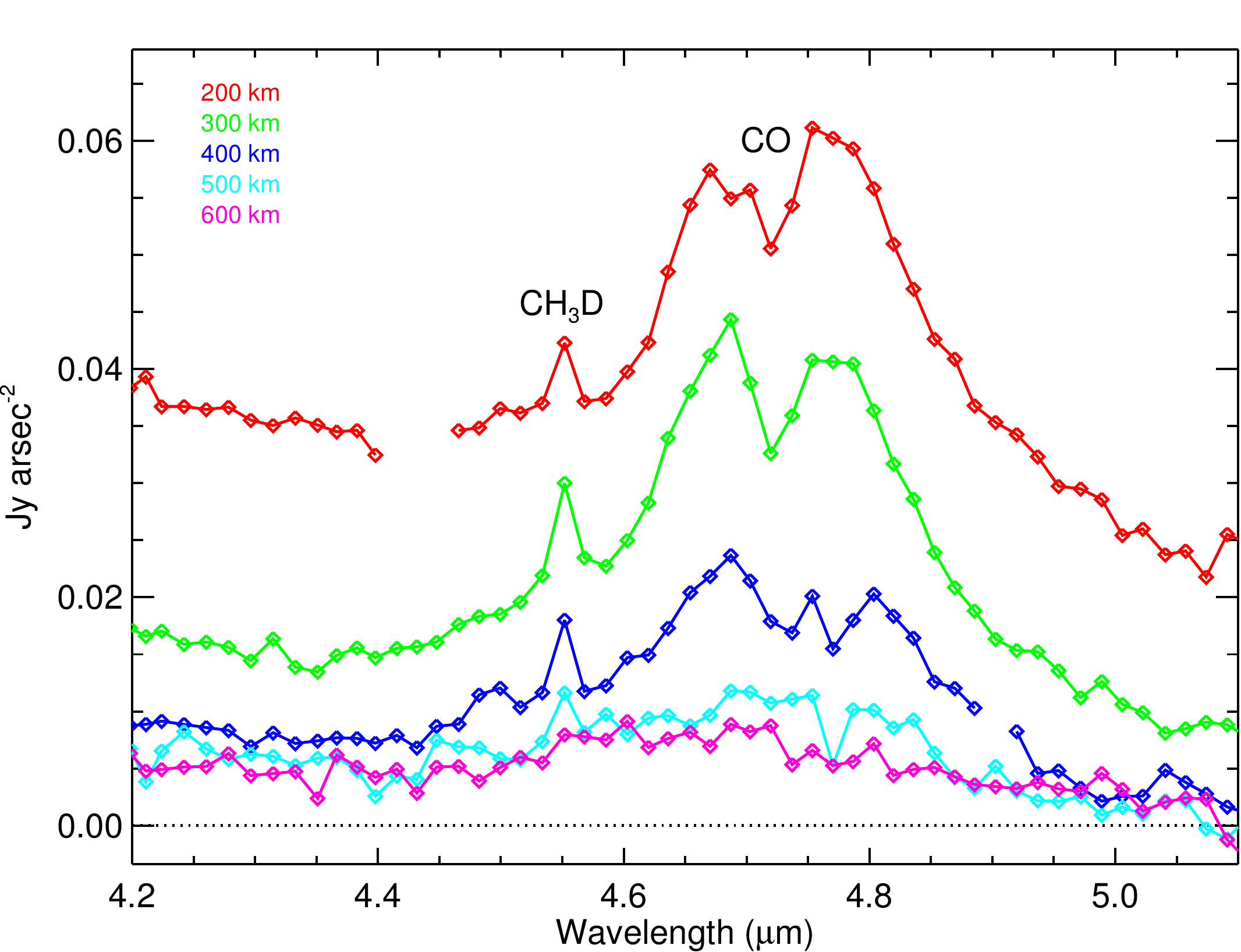} } 
\caption{VIMS limb spectra showing the CO and \dmethane\  spectral region. The scattering contribution at the lowermost altitudes is evident.}
\label{fig:vimslimb1}
\end{center}
\end{figure}

\begin{figure}[ht]
\begin{center}
\rotatebox{-270}{\includegraphics[width=11cm]{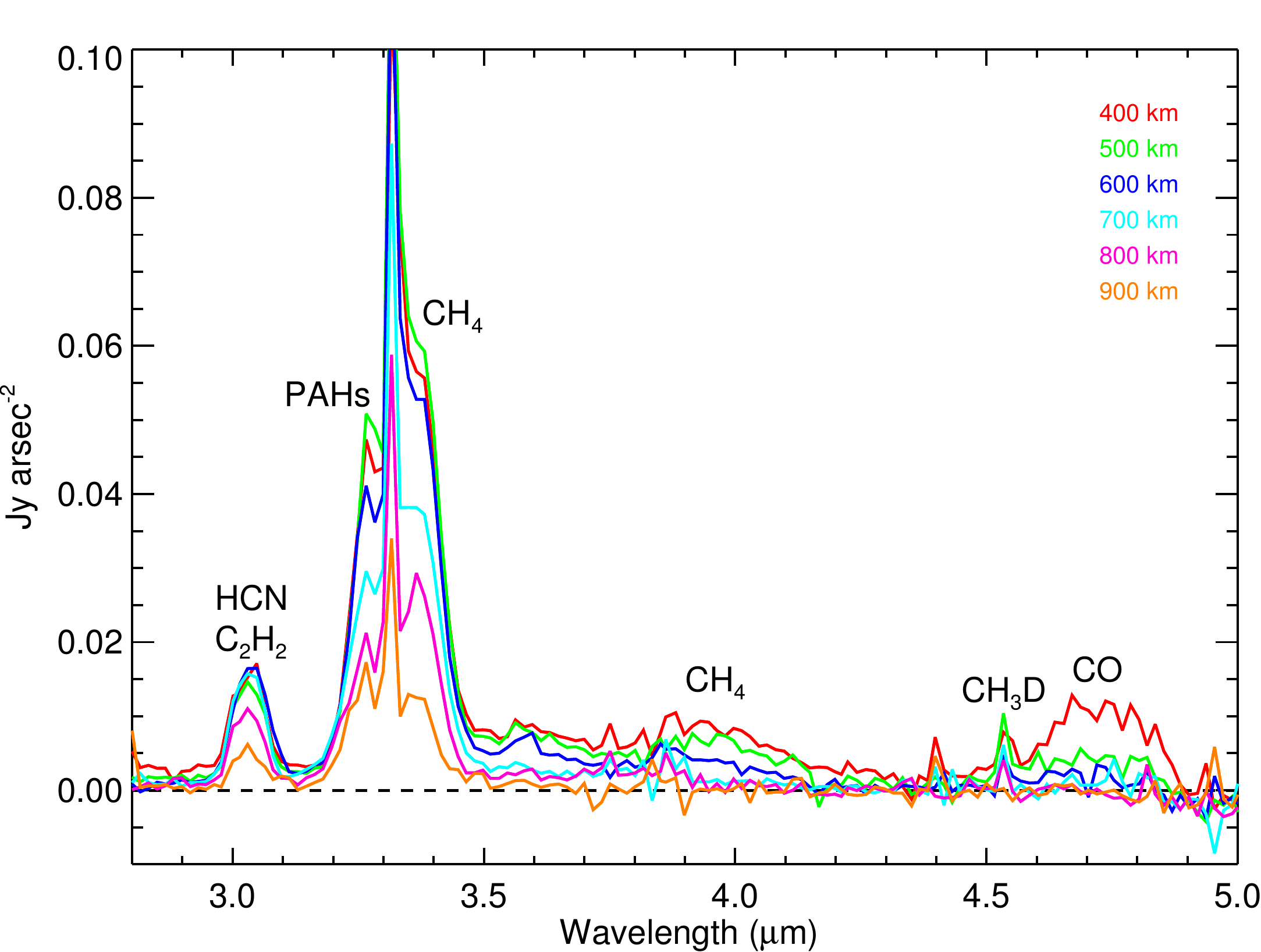} } 
\caption{Limb daytime spectra measured by VIMS of Titan middle/upper atmosphere during February and March 2009 (flybys T50 and T51) at tangent height from 400 up to 900 km. Vertical resolution is 30--40 km. Spectral resolution R$\sim$300 ($\sim$16 nm at 3.3\,$\mu$m). Spectral features of several species are shown. The PAHs emission is not clearly seen and is present only at 700--1200 km.}
\label{fig:vimslimb2}
\end{center}
\end{figure}

\begin{figure}[ht]
\begin{center}
{\includegraphics[width=14cm]{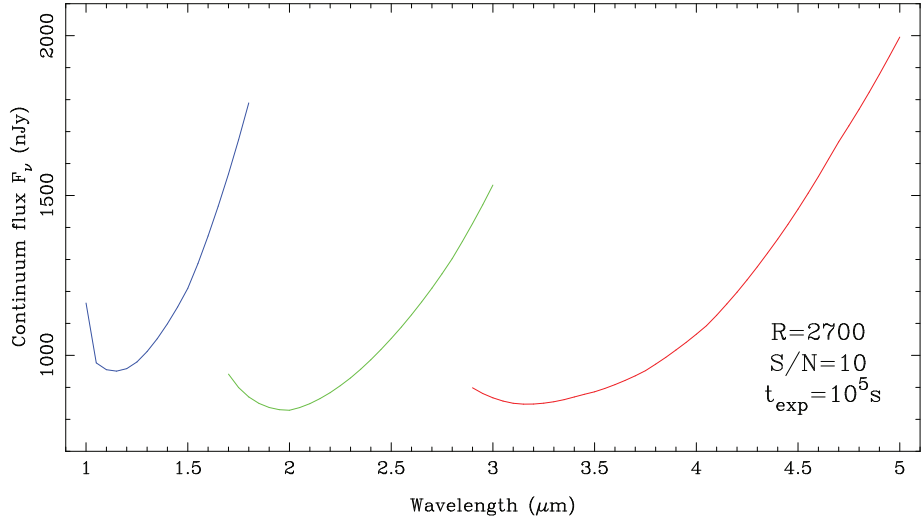} } 
\caption{NIRSpec sensitivity for S/N=10 and an exposure time of 10$^5$\,s.}
\label{fig:nirspec_sensi}
\end{center}
\end{figure}

\begin{figure} 
\plotone{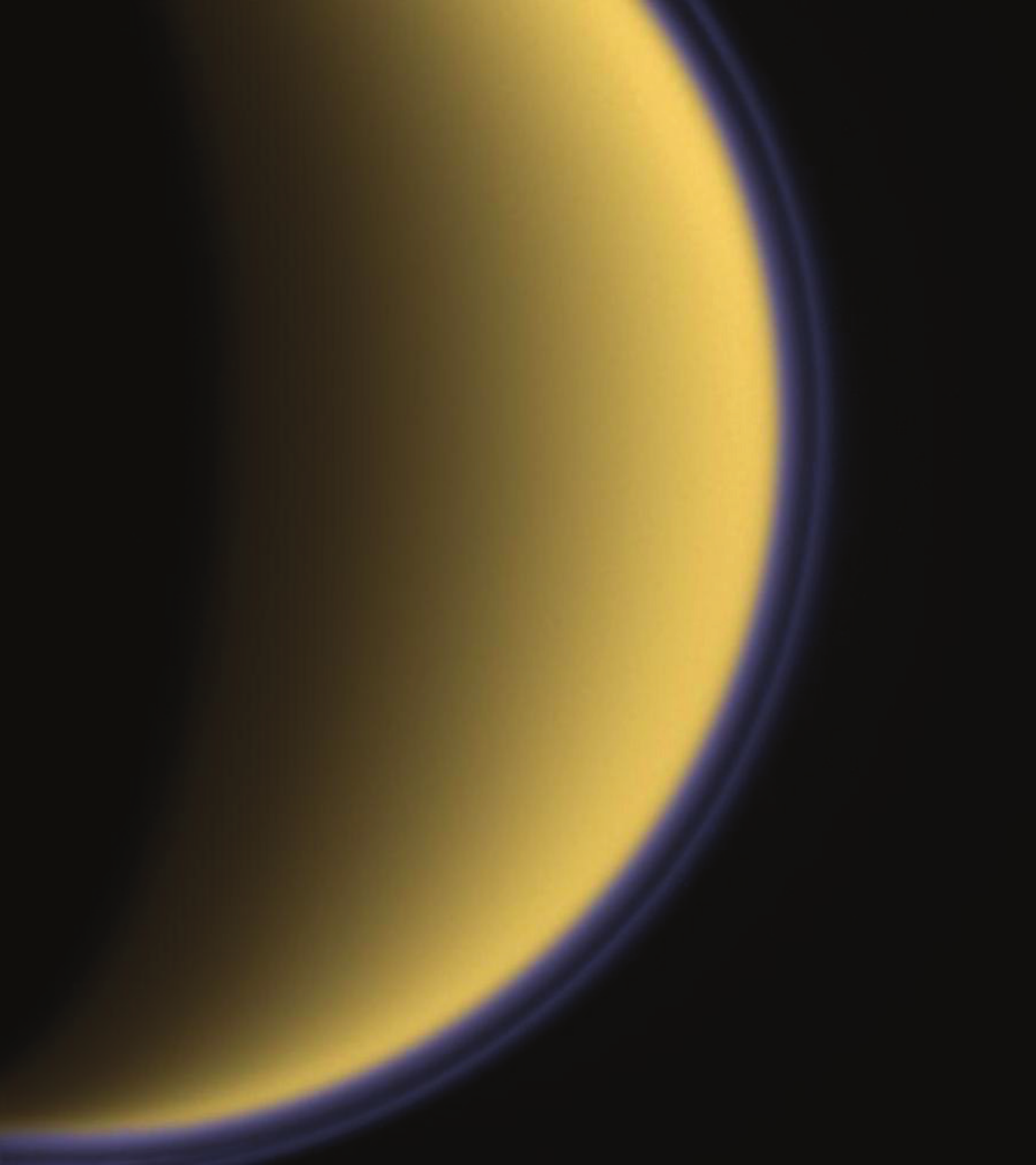}
\caption{Cassini image of Titan's southern hemisphere, processed to enhance UV (338 nm) wavelengths to show the striking detached haze layer floating above the main haze at an altitude of 500 km. Image PIA 06090, July 3rd 2004. Credit: NASA/JPL/Space Science Institute.}
\label{fig:haze}
\end{figure}

\begin{figure}
\plotone{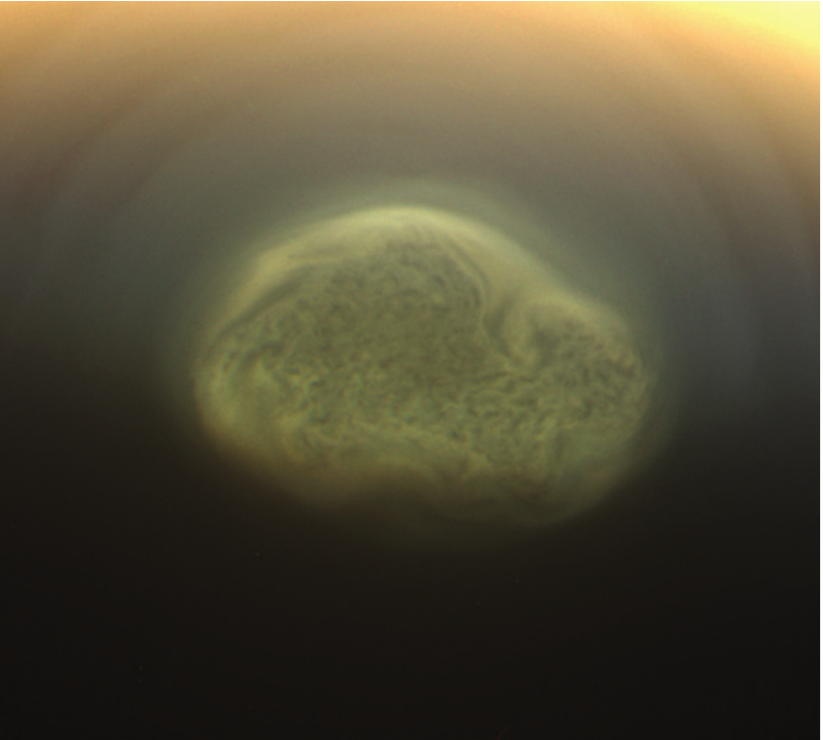}
\caption{Southern polar vortex cloud imaged by Cassini on June 27th 2012 at a range of 484,000 km \citep{west15}.  HCN ice/liquid has since been inferred to be a component of this cloud \citep{dekok14}. Image credit: NASA/JPL/Space Science Institute.}
\label{fig:spolecloud}
\end{figure}

The authors wish to express their thanks to Stefanie Milam, Dean Hines and John Stansberry of the Solar System Working Group (SSWG), and Pierre Ferruit (ESA/NIRSpec) for answering technical questions and giving helpful feedback during the writing of this paper. Don Jennings supplied the CIRS Titan spectrum (Fig.~\ref{fig:cirsspec}). A. Adriani,  M. L. Moriconi and B.M. Dinelli assisted by supplying the Cassini/VIMS limb data in Figs.~\ref{fig:vimslimb1} and \ref{fig:vimslimb2}. NAT is funded by the UK Science and Technology Facilities Council and the UK Space Agency. SR acknowledges financial support from the French ``Agence Nationale de la Recherche" (ANR Project: \methane @Titan and ANR project ``APOSTIC" \#11BS56002), France. TC is funded by the ESA Research Fellowship Programme in Space Science.  M. L.-P. was supported by the Spanish MCINN under grants AYA2011-23552 and ESP2014-54362-P. The authors are grateful to one anonymous reviewer for very helpful comments and feedback.



{\it Facilities:} \facility{Cassini (ISS, VIMS, CIRS)}, \facility{HST (WFPC-2)}, \facility{JWST (NIRSpec, NIRCam, MIRI)}.







\clearpage



\clearpage

\clearpage

\clearpage

\nocite{dekok14}







\clearpage

\begin{deluxetable}{lllllll}
\tabletypesize{\scriptsize}
\rotate
\tablecaption{Overview of JWST instrument specifications relevant to Titan science.\label{tab:specs}}
\tablewidth{0pt}
\tablehead{
\colhead{Instrument} & \colhead{Mode} & \colhead{Spectral Range} & \colhead{Resol. Power} & \colhead{FOV\tablenotemark{a}} &
\colhead{Pixel Pitch} & \colhead{Spatial Resol.\tablenotemark{b}}
}
\startdata
{\bf NIRSpec} & IFU\tablenotemark{c} & $0.6-5.0$~\micron & $R\sim$100, 1000, 2700 & 3{\arcsec}{$\times$}3{\arcsec} & 0.1\arcsec{$\times$}0.1\arcsec & 650 km/pix\\
 & SLIT & $0.6-5.0$~\micron & $R\sim$100, 1000, 2700 & 0.4{\arcsec}{$\times$}3.8{\arcsec} & & 650 km/pix \\
 & & & & & & \\
{\bf NIRCam} & SW\tablenotemark{d} & $0.6-2.3$~\micron & $R\sim$4 (W), 10 (M), 100 (N)\tablenotemark{e} & $2.2\times 2.2$\arcmin & 0.0317\arcsec & 200 km/pix \\ 
  & LW\tablenotemark{f} & $2.4-5.0$~\micron & $R\sim$4 (W), 10 (M), 100 (N)\tablenotemark{g} & $2.2\times 2.2$\arcmin & 0.0648\arcsec  & 400 km/pix \\ 
 & & & & & & \\
 {\bf NIRISS} & ES Spect.\tablenotemark{h} & $1.0-2.5$~\micron & $R\sim$150 & $2.2\times 2.2$\arcmin & 0.0654\arcsec & 400 km/pix \\ 
   & SS Spect.\tablenotemark{i} & $0.6-3.0$~\micron & $R\sim$700 & $2.2\times 2.2$\arcmin & 0.0654\arcsec & 400 km/pix \\ 
   & Imaging & $1.0-5.0$~\micron &  & $2.2\times 2.2$\arcmin & 0.0654\arcsec & 400 km/pix \\ 
 & & & & & & \\
 {\bf MIRI} & MRS\tablenotemark{j}  1A  & $4.96-7.71$~\micron & $R\sim$3250 & $3.00\times 3.87$\arcsec & $0.18{\times}0.19$\arcsec & 1200 km/pix \\  
                 & MRS\tablenotemark{j} 1B  & $7.71-11.90$~\micron & $R\sim$2650 & $3.50\times 4.42$\arcsec & $0.28{\times}0.19$\arcsec & 1800 km/pix \\  
                 & MRS\tablenotemark{j} 2A  & $11.90-18.35$~\micron & $R\sim$2000 & $5.20\times 6.19$\arcsec & $0.39{\times}0.24$\arcsec & 2500 km/pix \\  
                & MRS\tablenotemark{j} 2B  & $18.35-28.30$~\micron & $R\sim$1550 & $7.60\times 7.60$\arcsec & $0.64{\times}0.27$\arcsec & 4100 km/pix \\ 
\enddata
\tablecomments{Instrument specifications are current at time of publication. The reader is refered to on-line specifications at STScI for recent/final performance characteristics.}
\tablenotetext{a}{Field of View size.}
\tablenotetext{b}{Based on computed, diffraction-limited, Airy disk angular radius, and Titan angular diameter of 0.8\arcsec .}
\tablenotetext{c}{Integral Field Unit (Image Slicer).}
\tablenotetext{d}{Short Wavelength Imager.}
\tablenotetext{e}{Filters are: wide: F070W, F090W, F115W, F150W, F200W; medium: F140M, F162M, F182M, F210M; narrow: F164N, F187N, F212N.}
\tablenotetext{f}{Long Wavelength Imager.}
\tablenotetext{g}{Filters are: wide: F277W, F356W, F444W; medium: F250M, F300M, F335M, F360M, F410M, F430M, F460M, F480M; narrow: F323N, F405N, F466N, F470N.}
\tablenotetext{h}{Extended Source Spectroscopy mode.}
\tablenotetext{i}{Single-Source Spectroscopy mode.}
\tablenotetext{j}{Medium Resolution Spectrometer (Integral Field Mode).}
\end{deluxetable}

\clearpage

\begin{table}
\begin{center}
\caption{Titan cloud science goals with JWST.\label{tab:clouds}}
\begin{tabular}{lll}
\tableline\tableline
Timescale & Scientific Goals & Observations Required \\
\tableline
& & \\
 SINGLE  & Cloud detection and physical  & Joint detection/observation \\
OBSERVATION & characterization. & by NIRCam (direct imaging) and \\
 & & NIRSpec (spectral characterization). \\
  & & \\
HOURS & Real-time cloud evolution & May not be possible with JWST, since \\
 & & 48-hr response time may be too slow, \\
 & & unless observations can be targeted \\
  & & ahead during times of likely cloud activity. \\
  & & \\
 DAYS & Cloud size changes & Will be possible using TOO request. \\ 
  & and dissipation & NIRSpec and NIRCam both used, with \\ 
  & & NIRCam supporting NIRSpec target \\
  & &  acquisition. \\
   & & \\
   WEEKS  & Surface rainfall evaporation & NIRSpec can be used to monitor changes \\
    TO MONTHS & & in surface spectrum over time following \\
    & &  large cloud outbreaks. NIRCam can be used \\ 
    & & to monitor cloud activity and to detect and  \\ 
    & & map possible large-scale surface darkening \\ 
    & & (and potentially later brightening). \\
    & & \\
    SEASONAL	& Cloud locations and distributions;  & Regular (e.g. biweekly) short observations \\ 
    & comparison to GCM/precipitation  & with JWST NIRCam will be especially  \\
    & models & helpful to monitor seasonal changes in cloud \\
    & & apparitions. \\
\tableline
\end{tabular}
\end{center}
\end{table}






\end{document}